\documentclass{iopart}

\usepackage{epsfig, psfrag,rotating} 
\usepackage{iopams}
\usepackage{mytheorem}
\def\PRL{{\em Phys. Rev. Lett.} }
\def\PRA{{\em Phys. Rev.} A }
\def\PRB{{\em Phys. Rev.} B }

\def\PRE{{\em Phys. Rev.} E }
\def\PR{{\em Phys. Rev.} }
\def\ZPB{{\em Z. Phys.} B }

\def\EPL{{\em Europhys. Lett.} }
\def\JPA{{\em J. Phys.} A }
\def\JPC{{\em J. Phys.} C }
\def\JSP{{\em J. Stat. Phys.} }
\def\PA{{\em Physica} A }
\def\PD{{\em Physica} D }
\def\PTP{{\em Prog. Theor. Phys.} }
\def\ACP{{\em Adv. Chem. Phys.} }

\newcommand{\fnt}{\footnotesize}

\newcommand{\sst}{\scriptscriptstyle}
\newcommand{\pts}[1]{\mskip+#1\thinmuskip}
\newcommand{\nts}[1]{\mskip-#1\thinmuskip}

\newcommand{\arsinh}{\mathrm{arsinh}}

\newcommand{\artanh}{\mathrm{artanh}}

\newcommand{\N}{\mathbb{N}}
\newcommand{\R}{\mathbb{R}}
\newcommand{\Z}{\mathbb{Z}}

\newcommand{\supp}{\mathrm{supp} \;}
\newcommand{\dhu}{d_{\rm Hutch}}
\newcommand{\lip}{\mathrm{Lip}(f) \leq 1}
\newcommand{\dint}{\int\!\!\int}
\newcommand{\norm}[1]{|\!|#1|\!|}
\newcommand{\charact}[1]{{\bf 1}_{#1}}
\newcommand{\eps}{\varepsilon}
\newcommand{\prob}[2]{p^{(#1)}_{#2}}
\newcommand{\cantor}[2]{\mathcal{C}^{(#1)}_{#2}}
\newcommand{\limcan}[1]{\mathcal{C}_{#1}}
\newcommand{\msu}[2]{\mu^{(#1)}_{#2}}

\newcommand{\binom}[2]{ {#1 \choose #2} }
\newcommand{\trap}[2]{T^{(#1)}_{#2}}
\newcommand{\polylog}{\mathrm{Li}}
\newlength{\figwidth}
\eqnobysec

\begin{document}

\title[Convolution of multifractals and the local magnetization]{Convolution of multifractals and the local magnetization in a random
  field Ising chain} 
\author{Thomas Nowotny and Ulrich Behn}

\address{Institut f\"ur Theoretische Physik, Universit\"at Leipzig, \\ 
Augustusplatz 10, 04109 Leipzig, Germany}

\begin{abstract}
  The local magnetization in the one-dimensional random-field Ising model is
  essentially the sum of two effective fields with multifractal probability
  measure. The probability measure of the local magnetization is thus the
  convolution of two multifractals. In this paper we prove relations between
  the multifractal properties of two measures and the multifractal properties
  of their convolution. The pointwise dimension at the boundary of the
  support of the convolution is the sum of the pointwise dimensions at the
  boundary of the support of the convoluted measures and the generalized box
  dimensions of the convolution are bounded from above by the sum of the
  generalized box dimensions of the convoluted measures.  The generalized box
  dimensions of the convolution of Cantor sets with weights can be calculated
  analytically for certain parameter ranges and illustrate effects we also
  encounter in the case of the measure of the local magnetization.  Returning
  to the study of this measure we apply the general inequalities and present
  numerical approximations of the $D_q$-spectrum. For the first time we are
  able to obtain results on multifractal properties of a physical quantity in
  the one-dimensional random-field Ising model which in principle could be
  measured experimentally. The numerically generated probability densities
  for the local magnetization show impressively the gradual transition from a
  monomodal to a bimodal distribution for growing random field strength $h$.
\end{abstract}


\section{Introduction} \label{introsec}
Multifractal measures appear in a variety of contexts. The one-dimensional
random-field Ising models \cite{Bruinsma}-\cite{Nieuwenhuizen} and random
exchange \cite{Derrida}-\cite{Tanaka2} Ising models as well as other
one-dimensional disordered systems \cite{Schmidt}-\cite{Luckbuch}, Bernoulli
convolutions \cite{Ledrappier} and even learning in neural networks
\cite{vanHemmen}-\cite{Radons3} are prominent examples. The use of a
reduction scheme for the partition function of the one-dimensional
random-field Ising model first introduced by Ruj\'an \cite{Rujan} results in
the partition function of a one spin system in an effective field (one-sided
reduction, spin at the boundary) \cite{Gyoergyi1}-\cite{bvhklz} or in two
effective fields (two-sided reduction, spin in the bulk) \cite{Bleher} which
is the appropriate point of view when investigating the local magnetization.
The effective fields are governed by an iteration thus giving rise to a
random iterated function system which is known to have a unique invariant
measure \cite{Hutchinson}.  The invariant measure of the effective fields is
typically a multifractal \cite{Bruinsma, Gyoergyi1, Behn4}.

In a recent publication \cite{Nowotny} phase transitions in the
$D_q$-spectrum of this invariant measure of the effective field were
investigated (cf also \cite{Behn6, Patzlaff}) and tight
bounds on the $D_q$ based on the pointwise dimension at specific points
generalizing results in \cite{Evangelou} were
formulated. The combination of both allows a more or less complete
understanding of the $D_q$-spectrum of the invariant measure of the effective
field by exclusively analytical methods.  Naturally the question arises
whether these results are relevant for physical quantities such as
correlation functions or the local magnetization which in principle are
experimentally accessible.

The local magnetization can be expressed as a function of the effective
fields \cite{Gyoergyi1}-\cite{Behn4} and we show in this paper that a
considerable amount of the knowledge of the multifractal properties of the
invariant measure of the effective field can be transferred to the measure of
the local magnetization. Being essentially the sum of two effective fields
with multifractal probability measure the local magnetization has a
probability measure which is the convolution of two multifractal measures. We
therefore first prove general relations between the multifractal properties
of two measures and the multifractal properties of their convolution which
then can be applied to the random-field Ising chain.  As the convolution of
measures is the composition of constructing the product measure and
projecting it in a certain way, the work on projections of multifractal
(product) measures \cite{Radons0, Falconer} is related to our subject.
Whereas these papers focus mainly on properties of projected measures with
respect to typical projections we are here concerned with a given projection
leading to the convolution. This special case need not necessarily have the
properties of a generic projection.  There is also related work on the
superposition of multifractals \cite{Radons1}-\cite{Stoop1} and some remarks
on the convolution of multifractals in \cite{Torre}.

In addition to the mathematical results we also calculate the measure of the
local magnetization and its multifractal spectrum numerically. random-field
Ising systems can be realized as dilute antiferromagnets in uniform magnetic
fields \cite{Belanger} and the local magnetization can in principle be
measured by neutron scattering or M\"o{\ss}bauer spectroscopy. The
probability distribution of the local magnetization with respect to the
disorder therefore should be experimentally accessible and could be compared
to our numerical results presented in figure \ref{fig9}. Especially the
gradual transition from a strongly peaked monomodal distribution to a
strongly peaked bimodal distribution observed numerically should clearly be
visible. Depending on the quality of the measurement it is even feasible to
calculate the $D_q$-spectrum of the obtained probability distribution by the
box methods described in section \ref{numericsec} and to compare to the
results presented here. This should at least reproduce the general form of
the multifractal spectrum shown in figure \ref{fig7}.

The paper is organized as follows. After recalling the model and the
reduction scheme in section \ref{modelsec} we prove general bounds on the
$D_q$-spectrum of the convolution of two measures and relations between
certain pointwise dimensions in section \ref{generalsec}. The results are
applied to the situation of the local magnetization in the random-field Ising
chain.  We then explicitly calculate some $D_q$-spectra in the simplified
situation of the convolution of equal-scale Cantor sets with weights in
section \ref{cantorsec}. In section \ref{numericsec} we present numerical
results for the $D_q$-spectrum of the measure of the local magnetization and
in the concluding section \ref{concludesec} we summarize our results and draw
some conclusions.

\section{The model} \label{modelsec}
In the following we
consider the one-dimensional random-field Ising model with quenched disorder
which for $N$ spins has the Hamiltonian
\begin{equation}
H_N(\{s\}_N)= - J \sum_{i=a}^{b-1} s_is_{i+1} - \sum_{i=a}^{b} h_is_i,
\end{equation}
with $a < 0 < b$ and $b-a+1=N$. $s_i$ denotes the classical spins at site
$i$ of the chain taking values $\pm 1$, $J$ is the coupling strength between
spins and $h_i$ is the random field at site $i$. The random fields are
independent identically distributed random variables with probability density
\begin{equation}
  \rho (h_i)= \frac{1}{2} \delta (h_i-h)+\frac{1}{2} \delta (h_i+h), \quad h
  \in \R^+ . \label{hndist}
\end{equation}
In former work we used a reformulation of the canonical partition function $Z_N=
\sum_{\{s\}_n} \exp( -\beta H_N(\{s\}_N))$ to the partition function of the spin $s_a$ at the
left-hand boundary of the chain in an effective field $x_a^{(N)}$ which was first
introduced by Ruj\'an \cite{Rujan},
\begin{eqnarray}
  Z_N &= \sum_{s_a = \pm 1} \exp \Big(\beta\Big[x_a^{(N)} s_a + \sum_{i=a+1}^{b}
  \!\! B(x_i^{(N)})\Big]\Big) ,
  \\
  x_i^{(N)} &= A(x_{i+1}^{(N)}) + h_i , \quad x^{(N)}_{b+1}= 0
  \label{mapping} ,
\end{eqnarray}
with
\begin{eqnarray}
  A(x) &= (2\beta)^{-1} \ln (\cosh \beta (x+J)/\cosh \beta (x-J)) , \\
  B(x) &= (2\beta)^{-1} \ln (4 \cosh \beta (x+J)\cosh \beta (x-J)). 
\end{eqnarray}
When viewing (\ref{mapping}) as a random iterated function system (RIFS) we
will also write $x_n$ instead of $x_i^{(N)}$ for the effective field after
$n= N-i+1$ iterations of (\ref{mapping}). The effective fields
$x_n$ are random variables on the random-field probability space and we write
$p_n(x)$ for their induced probability density, $P_n(x) = \int_0^x p_n(\xi)
d\xi$ for their distribution function and $\mu_n (X) = \int_X p_n $ for their
measures. The iteration (\ref{mapping}) induces a Frobenius-Perron
(Chapman-Kolmogorov) equation for the distribution functions,
\begin{equation}
    P_{n}(x) = \int \! dh \; \rho (h)P_{n-1}\big(A^{-1}(x-h)\big) =
  \sum_{\sigma=\pm} \frac{1}{2} P_{n-1}\big(f_{\sigma}^{-1}(x)\big) ,
  \label{frobenius}
\end{equation}
and accordingly for the densities and measures. The symbols $f_\pm$ denote
the functions $f_\pm (x) := A(x) \pm h$. The Frobenius-Perron equation
has a unique invariant measure $\mu^{(x)}$ and the measures $\mu_n^{(x)}$
converge to $\mu^{(x)}$ in the weak topology of Borel measures on $\R$. The
invariant measure $\mu^{(x)}$ therefore is the measure of the effective field
$x$ in the thermodynamic limit $b \to \infty$ ($n \to \infty$ in the notation
of the RIFS).

In this paper we focus our interest on the local
magnetization in the bulk which is given by $m_{i,N}^{\rm bulk} = \langle s_i
\rangle_N$ at some site $a < i < b$ inside the chain. To obtain $\langle s_i
\rangle_N$ we rewrite the partition function to a one-spin partition function
with remaining spin $s_i$,
\begin{equation}
  \fl 
  Z_N= \sum_{s_i = \pm 1} \exp\left(\beta\left[\big(x_i^{(N)}
  +A(y_{i-1}^{(N)})\big) s_i
  \smash{+\sum_{j=a}^{i-1} B(y_j^{(N)}) + \! \sum_{j=i+1}^{b} \!
  B(x_j^{(N)})}\right]\right) \vphantom{\sum_{j=a}^{i-1}} ,
\end{equation}
with two effective fields
\begin{eqnarray}
  x_j^{(N)}= A(x_{j+1}^{(N)}) +h_j  , \quad i \leq j \leq b, \quad
  x_{b+1}^{(N)}= 0, \\ 
  y_j^{(N)}= A(y_{j-1}^{(N)}) +h_j  , \quad a \leq j < i, \quad y_{a-1}^{(N)}= 0 
\end{eqnarray}
from the right and the left of site $i$ respectively.
The local magnetization at $i$ is thus given by \cite{Gyoergyi1}-\cite{ Behn4}
\begin{equation}
  m_{i,N}^{\rm bulk}= \langle s_i \rangle_N = \tanh \beta \big( x_i^{(N)} +
  A(y_{i-1}^{(N)})\big) . \label{magnet}
\end{equation}
Let us introduce the notation
\begin{equation}
  f_\# (\mu) (X) := \mu(f^{-1}(X))
\end{equation}
for the mapping on Borel measures induced by a measurable
function $f$, e.g. $\tanh \beta_\# (\mu) (X) = \mu(1/\beta \, \artanh (X))$.
For the measure of $m_{i,N}^{\rm bulk}$ we obtain in this notation
\begin{equation}
    \mu^{(m)}_{l,r} = \tanh
\beta_\# (\mu^{(x)}_l \ast A_\# \mu^{(y)}_r) 
\end{equation}
with $l= i-a-1$ and $r= b-i$. As the effective fields share the same
Frobenius-Perron equation (\ref{frobenius}) and the invariant measure of this
equation is unique, the measures $\mu^{(x)}$ of the right-hand effective field in
the thermodynamic limit $b \to \infty$ and $\mu^{(y)}$ of the left-hand effective
field in the thermodynamic limit $a \to -\infty$ are identical. Therefore, as
we will see below, the measure $\mu^{(m)}_{l,r}$ of the local magnetization
in the bulk converges to
\begin{equation}
\mu^{(m)} = \tanh \beta_\# (\mu^{(x)} \ast A_\# \mu^{(x)}) \label{mumagnet}
\end{equation}
in the thermodynamic limit $a \to -\infty$, $b \to \infty$ (cf lemma
\ref{lemma1} below), i.e.~the local magnetization $m_{i,N}^{\rm bulk}$
converges in distribution to a random variable $m^{\rm bulk}$ with measure
$\mu^{(m)}$. Please note that $\mu^{(m)}$ is space independent because of the
uniqueness of the invariant measure of the Frobenius-Perron equation and the
continuity of the convolution.

The local magnetization {\em at the boundary} on the other hand is obtained
if we consider only one effective field, i.e.
\begin{equation}
  m^{\rm boundary}= \langle s_a \rangle = \tanh \beta x \label{mbound}
\end{equation}
which has the measure $\tanh \beta_\# \mu^{(x)}$. The multifractal properties of
$\mu^{(x)}$ are well known (cf \cite{Nowotny, Behn6}) and
general arguments show that $\tanh \beta_\#$ has no effect on the
$D_q$-spectrum (cf \cite{Riedi}) such that the results apply to the measure of
$m^{\rm boundary}$ as well. The main point of this paper is the
generalization to the magnetization of the bulk which is of greater physical interest.  

\section{Convolution of multifractals} \label{generalsec}
\begin{figure}
\begin{indented}
  \item
\psfrag{0}{$0$}
\psfrag{x}{$x$}
\psfrag{y}{$y$}
\psfrag{xm}{$x_-$}
\psfrag{ym}{$y_-$}
\psfrag{mu0}{$\mu \equiv 0$}
\psfrag{nu0}{$\nu \equiv 0$}
\psfrag{numu0}{$\mu \equiv 0$,}
\psfrag{numu1}{$\nu \equiv 0$}
\psfrag{e2}{$\frac{\eps}{2}$}
\psfrag{e}{$\eps$}
\psfrag{relevant}{\parbox{2cm}{\raggedright \footnotesize relevant region of
    integration}}
\figwidth=\textwidth
\addtolength{\figwidth}{-\mathindent}
\epsfig{file=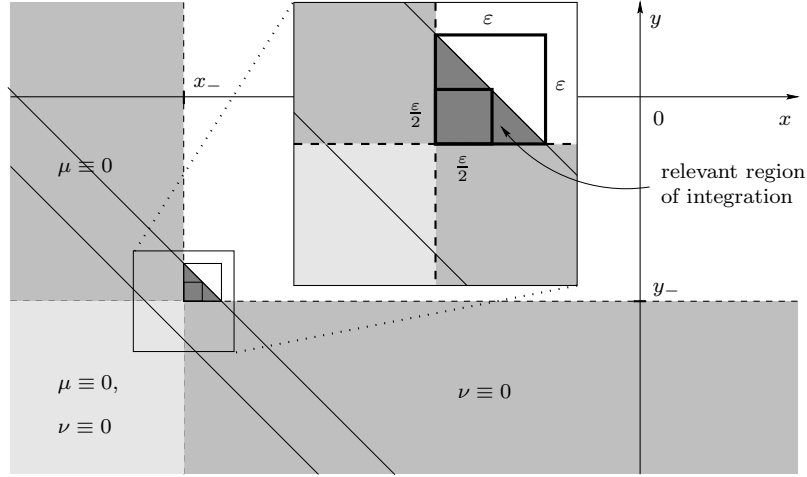,width=\figwidth}
\end{indented}
\caption{Illustration of the proof of lemma \ref{lemma2}. The diagonal strip
  is the region in which $\charact{B_\eps(z_-)}(x+y)$ is non-zero. Therefore,
  the dark grey triangle is the relevant region with non-zero contributions
  to the integral (\ref{mubeintegral}). As $\mu$ and $\nu$ are positive
  measures, integration over the small square of side length $\frac{\eps}{2}$
  provides a lower and integration over the larger square of side length
  $\eps$ an upper bound on the integral. \label{fig1}}
\end{figure}

In this section $(\mu_n)_{n \in \N}$ and $(\nu_n)_{n \in \N}$ denote
sequences of bounded Borel measures on $\R$ which are Cauchy sequences with
respect to the Hutchinson metric $\dhu$ (cf \cite{Hutchinson}). As the space
of bounded Borel measures on $\R$ is complete with respect to $\dhu$
\cite{Hutchinson}, $(\mu_n)$
and $(\nu_n)$ converge and we write $\mu:= \dhu$-$\lim_{n \to \infty} \mu_n$ and
$ \nu:= \dhu$-$\lim_{n \to \infty} \nu_n$.
As explained above we are interested in the properties of the convolution of
bounded Borel measures.
The convolution of two bounded Borel measures $\mu$ and $\nu$ is always well
defined (cf \cite{Bauer}) and will in the following be denoted by $\mu \ast \nu$.
As a first step we show that the convolution is
continuous with respect to $\dhu$.
\begin{lemma}[Continuity of the convolution] \label{lemma1} \mbox{} \\
  Let $(m_i)_{i\in \N}$ and $(n_i)_{i \in \N}$ be two monotonically growing
  unbounded sequences of natural numbers. Then $(\mu_{m_i} \ast
  \nu_{n_i})_{i \in \N}$ converges to a bounded Borel measure in Hutchinson
  topology and the limit is $\dhu$-$\lim_{i \to \infty} \mu_{m_i} \ast
  \nu_{n_i} = \mu \ast \nu$.
\end{lemma}
\begin{proof}
  Let $\eps > 0$. The convergence of $(\mu_n)_{n \in \N}$ and $(\nu_n)_{n
  \in \N}$ implies the existence of numbers $M, N \in \N$ such that for all $i
  \geq M$ $\dhu (\mu_{m_i} , \mu) \leq \eps$ and for all $i \geq N$
  $\dhu (\nu_{n_i}, \nu) \leq \eps$. Let $\tilde{N} := \max(M,N)$. For
  all $i \geq \tilde{N}$ we then have
  \begin{eqnarray}
    \fl
    \dhu(\mu_{m_i} \ast \nu_{n_i }, \mu \ast \nu) = \nonumber \\
\fl    \hphantom{=}\sup\left\{\int f(z) \,
      \mu_{m_i} \ast \nu_{n_i}(dz) -
      \int f(z) \, \mu \ast \nu(dz) \; \Big| \; \lip \right\}.
  \end{eqnarray}
  The definition of the convolution of two measures implies $\int f(z) \,
  \mu_{m_i} \ast \nu_{n_i}(dz) = \dint f(x+y) \mu_{m_i}(dx) \nu_{n_i}(dy)$
  and $\int f(z) \, \mu \ast \nu(dz) = \dint f(x+y) \mu(dx) \nu(dy)$.
  Inserting $0 = - \dint f(x+y) \mu(dx)\nu_{n_i}(dy) + \dint f(x+y)
  \nu_{n_i}(dy) \mu(dx) $ we obtain
  \begin{eqnarray}
    \fl
   \dhu(\mu_{m_i} \ast \nu_{n_i }, \mu \ast \nu) = \nonumber \\
    \fl  \hphantom{=}
    \eqalign{
    \sup \left\{\vphantom{\int}\right. &\int \left(
    \int f(x+y) \mu_{m_i}(dx) - \int f(x+y) \mu(dx)\right) \nu_{n_i}(dy) \\
   & + \left. \int \left( \int f(x+y) \nu_{n_i}(dy) - \int f(x+y) \nu(dy) \right)
    \mu(dx) \; \Big| \; \lip \right\}} \\
    \fl 
    \eqalign{
    \leq & \int \sup \left\{ \int f(x+y) \mu_{m_i}(dx) - \int f(x+y) \mu(dx)
   \; \Big| \; \lip \right\} \nu_{n_i}(dy) \\
     & +  \int \sup \left\{\int f(x+y) \nu_{n_i}(dy) - \int f(x+y) \nu(dy)
   \;\Big| \; \lip \right\} \mu(dx) . }
  \end{eqnarray}
As the condition $\lip$ is translationally invariant we further obtain
  \begin{eqnarray}
    \fl
     \sup\left\{ \int f(x+y) \mu_{m_i} (dx) - \int f(x+y) \mu(dx) \;\Big|\; \lip
    \right\} = \nonumber \\ \fl
    \hphantom{=} \sup \left\{ \int f(x) \mu_{m_i}(dx) - \int f(x) \mu(dx)
      \;\Big|\; \lip
    \right\}  = \dhu(\mu_{m_i}, \mu) \leq \eps.
  \end{eqnarray}
  In the same way
  \begin{eqnarray}
    \fl
    & \sup\left\{ \int f(x+y) \nu_{n_i}(dy) - \int f(x+y) \nu(dy) \;\Big|\; \lip
    \right\} = \dhu(\nu_{n_i}, \nu) \leq \eps .
  \end{eqnarray}
  We thus arrive at
  \begin{eqnarray}
    \fl
    \dhu(\mu_{m_i} \ast \nu_{n_i }, \mu \ast \nu)  \leq \int \eps \,
  \nu_{n_i}(dy) + \int \eps \, \mu(dx) = (\norm{\nu_{n_i}} + \norm{\mu})
  \; \eps
  \end{eqnarray}
  in which $\norm{\nu_{n_i}}= \nu_{n_i}(\R)$ and $\norm{\mu}= \mu(\R)$ denote
  the total mass of $\nu_{n_i}$ and $\mu$ respectively.
  \rule{2cm}{0cm}
\end{proof}
As the metric $\dhu$ topology and the weak topology coincide on bounded Borel
measures with compact support \cite{Hutchinson} we have the following corollary.
\begin{corollary} \label{corollary1}
  If $\supp \mu$ and $\supp \nu$ are compact, ${\rm w}$-$\lim_{i \to \infty}
  \mu_{m_i} \ast \nu_{n_i} = \mu \ast \nu$. Furthermore $\supp \mu \ast \nu$
  is also compact.
\end{corollary}
For the situation of the two-sided random-field Ising chain considered in
this paper lemma \ref{lemma1} implies that the thermodynamic limit $l,r \to
\infty$ can be carried out in an arbitrary way and that the result is the
same as when first taking the thermodynamic limit for the effective fields
and then calculating the measure of the local magnetization. Having detailed
knowledge of the properties of the $D_q$-spectrum and the pointwise
dimensions of the invariant measure of the effective field it is now
interesting to investigate the relationship of the $D_q$-spectra and
pointwise dimensions of $\mu$ and $\nu$ to the $D_q$-spectrum and pointwise
dimensions of $\mu \ast \nu$. The following lemmata allow us to transfer the
knowledge about the multifractal properties of the invariant measure of the
effective field gathered in \cite{Nowotny} to the measure of the
local magnetization.
In the following we consider only measures with bounded, i.e.~compact support.
Let $x_-:= \min \, \supp \!\mu > -\infty$ denote the left and $x_+ := \max \,
\supp \! \mu
< \infty$ the right boundary of $\supp \mu$. For the boundaries of $\supp
\nu$ we write $y_-$ and $y_+$. The pointwise dimension at the boundary of the
support of $\mu \ast \nu$ can be obtained from the pointwise dimensions at
$x_+$, $x_-$, $y_+$ and $y_-$. 
\begin{lemma}[Pointwise dimension of $\bmu \ast \bnu$ at the boundary of its
  support] \mbox{} \label{lemma2} \\
  The left boundary of $\mu \ast \nu$ is $z_- = x_- + y_-$ and the pointwise
  dimension $D_p(z_-;\mu \ast \nu) = D_p(x_-; \mu) + D_p(y_-; \nu)$.
  The result for the right boundary is analogous.
\end{lemma}
\begin{proof}
  The pointwise dimension of $\mu \ast \nu$ at $z_-$ is defined as
  \begin{equation}
    D_p(z_- ; \mu \ast \nu) = \lim_{\eps \to 0} \frac{\ln(\mu \ast
    \nu(B_\eps(z_-)))}{\ln \eps} 
  \end{equation}
  in which $\mu\ast\nu (B_\eps(z_-))$ is given by
  \begin{equation}
    \mu\ast\nu (B_\eps(z_-)) = \dint \charact{B_\eps(z_-)}(x+y) \mu(dx)
    \nu(dy) .
    \label{mubeintegral}
  \end{equation}
  The symbol $\charact{X}$ denotes the characteristic function of a set $X$,
  i.e.~$\charact{X}(x) = 1$ if $x \in X$ and $=0$ otherwise.  The area in
  which $\charact{B_\eps(z_-)}(x+y)$ is non-zero is shown in figure
  \ref{fig1}. Neglecting the regions in which either $\mu = 0$ or $\nu = 0$
  (or both), the relevant region is the dark gray triangle. As $\mu$ and
  $\nu$ are positive measures, integration over the small square gives a
  lower and integration over the larger square an upper bound:
  \begin{equation}
    \fl
    \int_{y_-}^{y_-+\frac{\eps}{2}} \nts{5} \nu(dy) \int_{x_-}^{x_-
    +\frac{\eps}{2}} \nts{5}
    \mu(dx) \leq \mu\ast\nu (B_\eps(z_-)) \; \leq \; \int_{y_-}^{y_- +\eps}
    \nts{5} \nu(dy)
    \int_{x_-}^{x_- +\eps} \nts{5} \mu(dx) .
  \end{equation}
  Taking into account that $\mu = 0$ on $(x_- - \eps, x_-)$ and $\nu= 0$ on
  $(y_- -\eps, y_-)$ we can write
  \begin{equation}
    \fl
    \nu(B_{\frac{\eps}{2}}(y_-)) \mu(B_{\frac{\eps}{2}}(x_-)) \leq \mu \ast
    \nu(B_\eps(z_-)) \leq \nu(B_\eps(y_-)) \mu(B_\eps(x_-))
  \end{equation}
  to finally obtain
  \begin{eqnarray}
    \fl
    \frac{\ln \nu(B_{\frac{\eps}{2}}(y_-)) + \ln
    \mu(B_{\frac{\eps}{2}}(x_-))}{\ln \frac{\eps}{2} + \ln 2} \nonumber
  &\geq \frac{\ln \mu\ast\nu (B_\eps(z_-))}{\ln \eps}  \\ 
  &\geq \frac{\ln \nu(B_\eps(y_-)) + \ln \mu(B_\eps(x_-))}{\ln \eps} 
  \end{eqnarray}
  which completes the proof as both sides of the inequality converge to
  $D_p(x_-; \mu) + D_p(y_-; \nu)$ as $\eps \to 0$. The proof for the right
  boundaries is obtained by applying the same arguments to $\tilde{\mu}(X) :=
  \mu (-X)$ and $\tilde{\nu} (X) := \nu(-X)$.
\end{proof}

%%% Local Variables: 
%%% mode: latex
%%% TeX-master: "article"
%%% End: 
 To apply lemma \ref{lemma2} to the measure of the local
magnetization in the 1D RFIM it is important to know how the mappings $A_\#$
and $\tanh \beta_\#$ in (\ref{mumagnet}) influence the pointwise dimensions
of $\mu^{(m)}$. It turns out that they are of no significance in this context
because the pointwise dimension of the image measure at the image of some
point is the pointwise dimension of the original measure at this point if the
map under consideration is bi-Lipschitz.
\begin{lemma}[Stability of $\mathbf D_p$ with respect to bi-Lipschitz maps]
  \mbox{} \\
  Let $f: \R \to \R$ be a bi-Lipschitz function and $\mu$ a bounded Borel
  measure on $\R$. Then $D_p(f(x); f_\#(\mu))= D_p(x; \mu)$. \label{lemma3}
\end{lemma}
\begin{proof}
As $f$ is bi-Lipschitz so is $f^{-1}$ and therefore 
  \begin{equation}
    L^{-1} |y-x| \leq |f^{-1}(y) -f^{-1}(x)| \leq L |y-x|
  \end{equation}
  for some constant $L > 1$. Then
  \begin{equation}
    \fl
    |f^{-1}(f(x)+\eps) - x| = |f^{-1}(f(x)+\eps) - f^{-1}(f(x))| \leq L
     |f(x)+ \eps - f(x)| = L\eps
   \end{equation}
   and
   \begin{equation}
     |x-f^{-1}(f(x)-\eps)| \leq L|f(x) -(f(x) -\eps)| = L\eps .
   \end{equation}
   This implies $f^{-1}(B_\eps(f(x))) \subseteq B_{L\eps}(x)$. In the same
   way one obtains $B_{L^{-1} \eps} (x) \subseteq f^{-1}(B_\eps(f(x)))$ such
   that
   \begin{equation}
     \frac{\ln \mu(B_{L^{-1} \eps} (x))}{\ln L^{-1} \eps + \ln L}
     \geq \frac{\ln f_\# \mu (B_\eps(f(x)))}{\ln \eps}
     \geq \frac{\ln \mu(B_{L\eps}(x))}{\ln L\eps -\ln L}.
   \end{equation}
   The left and the right hand side of the
   inequality converge to $D_p(x; \mu)$ such that the middle part which converges
   to $D_p(f(x); f_\# \mu)$ also converges to this limit. 
\end{proof}
\begin{figure}
\psfrag{2i-2}{$\scriptstyle x_{2i-2}$}
\psfrag{2i}{$\scriptstyle x_{2i}$}
\psfrag{2i+2}{$\scriptstyle x_{2i+2}$}
\psfrag{x}{$x$}
\psfrag{y}{$y$}
\psfrag{e2}{$\scriptstyle \frac{\eps}{2}$}
\psfrag{e}{$\scriptstyle \eps$}
\psfrag{0}{$\scriptstyle 0$}
a)\parbox[t]{0.47\textwidth}{\rule{0.47\textwidth}{0cm}
  \epsfig{file=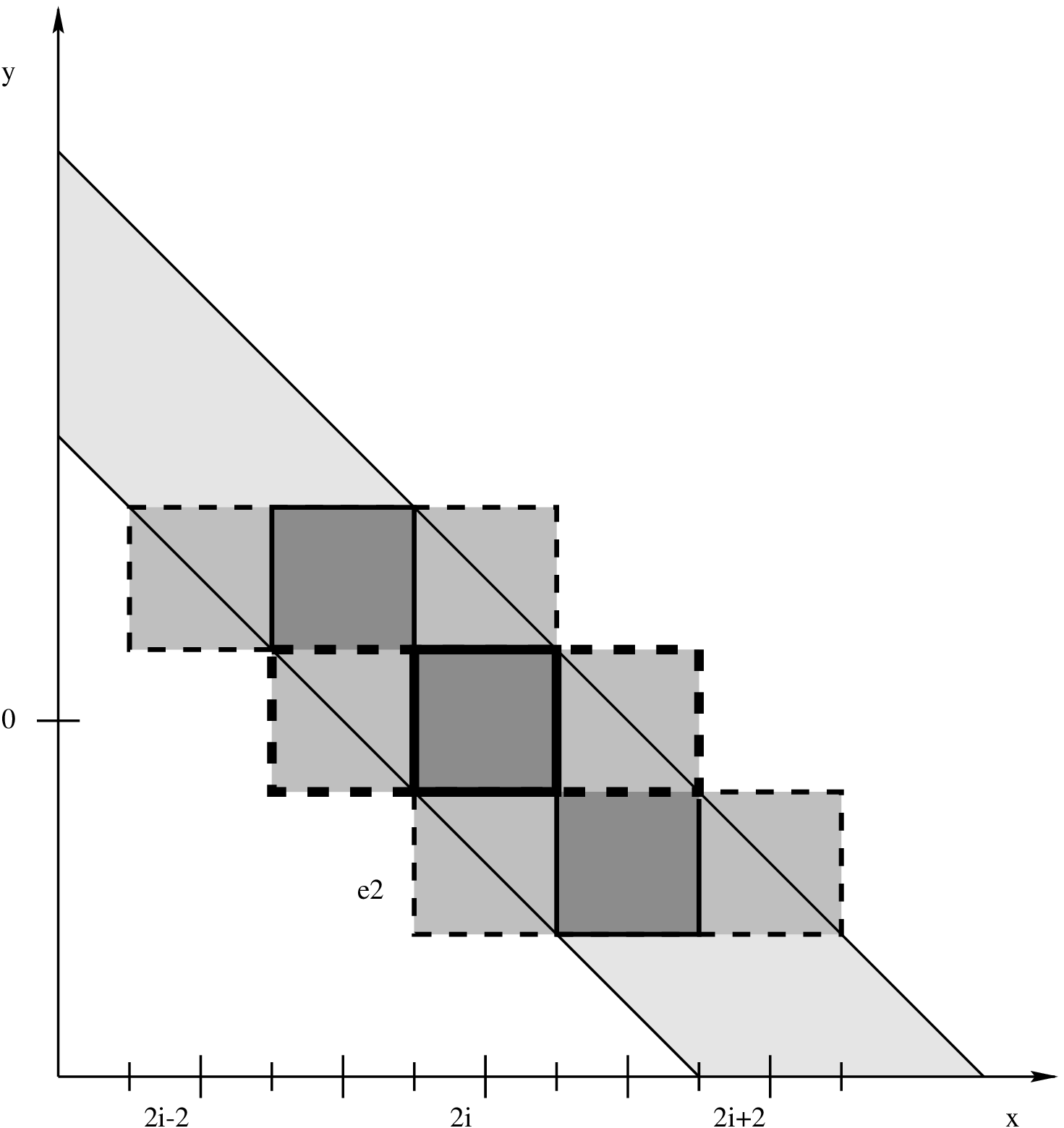, width=0.47\textwidth}} \hfill
b)\parbox[t]{0.47\textwidth}{\rule{0.47\textwidth}{0cm}
  \epsfig{file=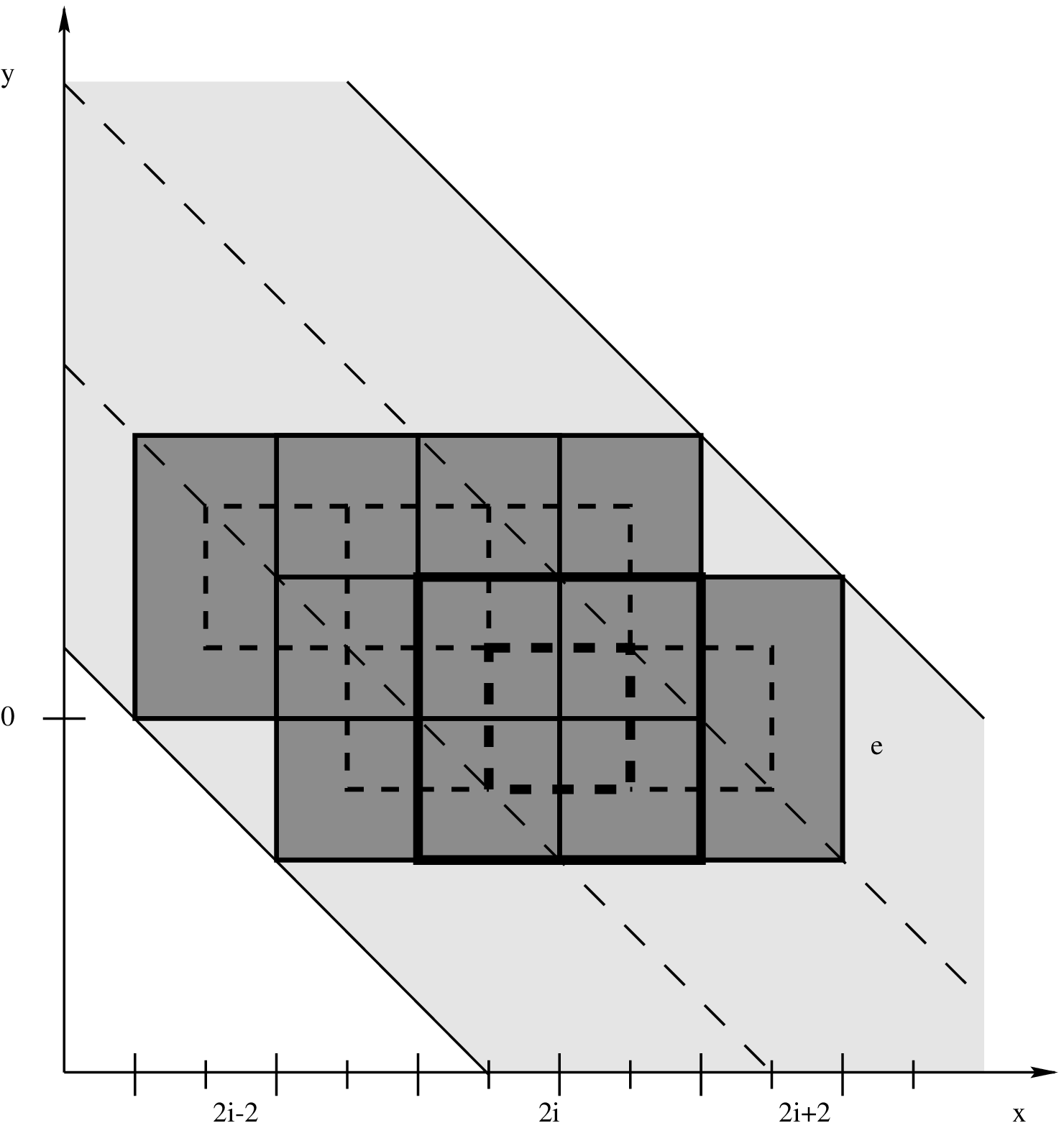, width=0.47\textwidth}}
\caption{Illustration of the main ideas of the proof of theorem
  \ref{theorem1}. Figure a) applies to $q>0$ and figure b) to $q<0$. The
  diagonal strip in a) represents the region of integration for $(\mu \ast
  \nu)_i$, the measure of one of the disjoint intervals of length $\eps$
  covering $\supp \mu \ast \nu$. The integral over the dark grey squares (and
  diagonally translated disjoint copies) provides a lower bound on $(\mu \ast
  \nu)_i$ used in the case $q > 1$. Considering additionally the integral
  over the dashed squares gives an
  upper bound on $(\mu \ast \nu)_i$ needed in the case $0 < q < 1$. \\
  The wide diagonal strip in b) is the region of integration for $(\mu \ast
  \nu)_i$, the measure of one of the (intersecting) enlarged intervals of
  length $3\eps$ covering $\supp \mu \ast \nu$.  The narrow dashed strip is
  the region of integration for the corresponding inner interval of size
  $\eps$. Integration over each of the six overlapping large squares of side
  length $\eps$ (solid lines) and their disjoint
  by $(-n \eps, n \eps)$ diagonally translated copies gives a lower bound on
  $(\mu \ast \nu)_i$ such that the sum of the six integrals gives a lower
  bound on $6 (\mu \ast \nu)_i$. The narrow strip is contained in the union
  of all the interior small squares of side length $\frac{\eps}{2}$ (dashed
  lines) assuring that the lower bound obtained is
  non-zero whenever the integral over the narrow strip is. This is an
  important point in the proof. The details are given in the text.
 \label{fig2}}
\end{figure}

%%% Local Variables: 
%%% mode: latex
%%% TeX-master: "article"
%%% End: 
 For bounded measures with compact support it is sufficient that
the function $f$ is bi-Lipschitz on an interval containing the support of the
measure. As $A(\cdot)$ and $\tanh \beta (\cdot)$ are bi-Lipschitz on any
finite interval lemma \ref{lemma2} and lemma \ref{lemma3} directly imply
$D_p(m_-; \mu^{(m)}) = D_p(m_-; \tanh \beta_\# (\mu^{(x)} \ast A_\#
\mu^{(x)})) = D_p(x_-; \mu^{(x)}) + D_p(A(x_-); A_\#\mu^{(x)}) = 2 D_p(x_-;
\mu^{(x)})$. In \cite{Nowotny} lower (upper) bounds on $D_q$ for
$q <0$ ($q>0$) based on the pointwise dimension at arbitrary points in the
support of the measure were developed. These bounds can directly be applied
to the $D_q$-spectrum of $\mu^{(m)}$ resulting in
\begin{equation}
  D_q(\mu^{(m)}) \geq \frac{q}{q-1} 2 D_p(x_-) = \frac{q}{1-q} \frac{2\ln
  2}{\ln A'(x_-)} \quad (q < 0).
\end{equation}
This bound is a tight bound as long as the pointwise dimension at the
boundary is weak. This is the case as long as $D_p(m_-) > 1$. The critical
value $h_c^{(m,3)}$ determined by this condition is
\begin{eqnarray}
  \frac{1}{2\beta} \ln\left(\frac{R+ e^{2\beta J}}{R^{-1} + e^{2\beta J}}
  \right) \quad {\rm with} \\
  R= 3\sinh (2\beta J) - e^{-2\beta J} +
  \sqrt{\big(3\sinh(2\beta J)-e^{-2\beta J}\big)^2 -1}.
\end{eqnarray}
The critical value can also be interpreted in terms of the measure
density. At this value of $h$ the measure density at the boundary of the
support changes from $0$ (for $ h < h_c^{(m,3)}$) to $\infty$ (for $h >
h_c^{(m,3)}$).

For $q > 0$ the corresponding bound is not of interest as the smoothness of
$p^{(x)}$ in the region of small $h$ implies smoothness of $p^{(m)}$ in this
region and thus $D_q = 1$ for all $q > 0$. The same applies to the
connectedness of the support implying $D_0=1$.  

The formerly discussed \cite{Nowotny, Behn6} transition in
the density of the effective field in which the slope of the coarse grained
measure density at the boundary of the support changes from $0$ to $\infty$
also has an analogue. This effect occurs for the coarse grained measure
density of the magnetization at $D_p(m_-) = \frac{1}{2}$ corresponding to
\begin{equation}
    h_c^{(m,4)} = \frac{1}{\beta} \arsinh\big(2^{-\frac{3}{2}} (1-9e^{-4\beta
J})^{\frac{1}{2}}\big) = h_c^{(3)} .
\end{equation}
Note that the measure density of the local magnetization changes its slope at
the boundary at
the same critical value at which the measure density of the effective field at the
boundary changes from $0$ to $\infty$, cf \cite{Nowotny, Behn6}.

Apart from the relation between the pointwise dimensions of the convolution
and its factors discussed so far there also exists a general relation
between the $D_q$-spectra. 
\begin{theorem}[Upper bound on $\mathbf D_q(\bmu \ast \bnu)$] \label{theorem1}
\mbox{} \\
  The $D_q$ spectrum of the convolution is bounded from above by the sum of
  the $D_q$-spectra of the factors,
  \begin{equation}
    D_q(\mu \ast \nu) \leq D_q(\mu) + D_q (\nu).
  \end{equation}
\end{theorem}
\begin{proof}
  We need to distinguish three cases, $q > 1$, $0 < q < 1$ and $ q < 0$.  For
  the first two cases the improved multifractal formalism with enlarged boxes
  coincides with the usual one and for simplicity we will use the later in
  these cases. Throughout the proof sums of the form $\sum_i \mu_i^q$ extend
  over all $i \in \Z$ with $\mu_i > 0$, i.e.~boxes with zero measure are not
  taken into account (which is important for $q \leq 0$). Let $\eps > 0$. We
  denote $x_i := \frac{\eps}{2} i$, $i \in \Z$. \\
  Let $q > 1$. For any $i \in \Z$
  \begin{equation}
    (\mu \ast \nu)_i := \mu \ast \nu(B_{\frac{\eps}{2}} (x_{2i}))
    = \dint \charact{B_{\frac{\eps}{2}} (x_{2i})}(x+y) \mu(dx) \nu(dy)
  \end{equation}
  is the integral over the diagonal strip in the $(x,y)$-plane shown in
  figure \ref{fig2}(a).
  Integration over the dark gray squares provides a lower bound on this
  integral.
  \begin{equation}
    (\mu \ast \nu)_i \geq \sum_j
    \mu(B_{\frac{\eps}{4}}(x_{2i+j}))\nu(B_{\frac{\eps}{4}}(x_{2i-j})) .
  \end{equation}
  Taking the $q$-th power of both sides and using $(\sum_i x_i)^q \geq \sum_i
  x_i^q$ for $q > 1$ and any positive numbers $x_i$ we obtain
  \begin{equation}
    (\mu \ast \nu)_i^q \geq
    \sum_j \mu(B_{\frac{\eps}{4}}(x_{2i+j}))^q
      \nu(B_{\frac{\eps}{4}}(x_{2i-j}))^q .
    \label{ineq1}
  \end{equation}
  For ${(\mu \ast \nu)'_i}^q :=
  \mu\ast\nu(B_{\frac{\eps}{2}}(x_{2i+1}))^q$ we have in the same way
  \begin{equation}
    {(\mu \ast \nu)'_i}^q \geq
    \sum_j \mu(B_{\frac{\eps}{4}}(x_{2i+j+1}))^q
    \nu(B_{\frac{\eps}{4}}(x_{2i-j}))^q .
    \label{ineq2}
  \end{equation}
  We denote $\mu_i := \mu(B_{\frac{\eps}{4}}(x_i))$ and $\nu_j:=
  \nu(B_{\frac{\eps}{4}}(x_j))$. Summing (\ref{ineq1}) and (\ref{ineq2}) and
  over all $i$ we get on the right hand side $\sum_i \sum_j \mu_i \nu_j$.  It
  is straightforward to show that $\sum_i {(\mu \ast \nu)'_i}^q \leq 2^{q+1}
  \sum_i (\mu\ast \nu)_i^q$ (cf \ref{appe1}) such that the left
  hand side of the sum of (\ref{ineq1}) and (\ref{ineq2}) summed over all $i$
  is less than or equal to $(2^{q+1}+1) \sum_i (\mu \ast \nu)_i^q$. Taking the
  logarithm, dividing by $\ln \eps$ and multiplying with $1/(q-1)$ we obtain
  \begin{equation}
    \frac{1}{q-1} \frac{\ln \sum_i (\mu \ast \nu)_i^q + \ln 2^{q+1}}
    {\ln \eps} \leq
    \frac{1}{q-1} \frac{\ln \sum_i \mu_i^q + \ln \sum_j \nu_j^q}
    {\ln \frac{\eps}{2} + \ln 2}
  \end{equation}
  which completes the proof for $q > 1$ as the left hand side converges to
  $D_q(\mu\ast\nu)$ and the right hand side to $D_q(\mu) + D_q(\nu)$ as $\eps
  \to 0$. \\
  Let $0 < q < 1$ and $i \in \Z$. We again write $(\mu \ast \nu)_i := \mu \ast
  \nu(B_{\frac{\eps}{2}} (x_{2i}))$, $\mu_i:= \mu(B_{\frac{\eps}{4}}(x_i))$
  and $\nu_j:= \nu(B_{\frac{\eps}{4}}(x_j))$. The solid and dashed squares
  in figure \ref{fig2}(a) and by $(n\eps, -n\eps)$ diagonally translated
  disjoint copies cover the diagonal strip over which we
  need to integrate to obtain $(\mu \ast \nu)_i$. We therefore have the upper
  bound 
  \begin{equation}
    (\mu \ast \nu)_i
    \leq \sum_j \sum_{k=-1}^{1} \mu_{2i+j+k}\nu_{2i-j} .
  \end{equation}
  Taking the $q$-th power and
  using $(\sum_i x_i)^q \leq \sum_i x_i^q$ for $q < 1$ and arbitrary positive
  numbers $x_i$ yields
  \begin{equation}
    (\mu \ast \nu)_i^q \leq \sum_j \sum_{k=-1}^{1} \mu_{2i+j+k}^q \nu_{2i-j}^q .
  \end{equation}
  When summing over all $i$ each combination $\mu_i \nu_j$ appears at most
  twice in the sum on the right hand side such that
  \begin{equation}
    \sum_i (\mu \ast \nu)_i^q \leq 2\sum_i \sum_j \mu_i^q \nu_j^q .
  \end{equation}
  Taking the logarithm of both sides,
  dividing by $\ln \eps$ and multiplying with $1/(q-1)$ results in
  \begin{equation}
    \frac{1}{q-1} \frac{\ln \sum_i (\mu\ast\nu)_i^q}{\ln \eps} \leq
    \frac{1}{q-1} \frac{\ln \sum_i \mu_i^q + \ln \sum_j \nu_j^q + \ln 2}
    {\ln \frac{\eps}{2} + \ln 2}.
  \end{equation}
  The limit $\eps \to 0$ yields $D_q(\mu \ast \nu) \leq D_q(\mu) +
  D_q(\nu)$. \\
  Let $q < 0$. In this case we need the improved multifractal formalism with
  enlarged intervals. We use the notation
  \begin{equation}
    (\mu\ast\nu)_i := \left\{\begin{array}{ll}
      \mu(B_{\frac{3}{2}\eps}(x_{2i}) &
      (\mu\ast\nu(B_{\frac{\eps}{2}}(x_{2i})) > 0) \\
      0 & ({\rm otherwise})
    \end{array} \right. .
  \end{equation}
  By this choice we enlarge the $\eps$-intervals by $\eps$ on both sides
  corresponding to $\kappa = 1$ in Riedi's notation \cite{Riedi}. Furthermore
  we denote
  \begin{equation}
    \fl
    \mu_i := \left\{ \begin{array}{ll}
       \! \mu(B_\frac{\eps}{2}(x_i)) & (\mu(B_{\frac{\eps}{4}}(x_i)) \geq 0)
        \\
       \! 0 & ({\rm otherwise})
      \end{array} \right. \nts{5}, \quad
    \nu_i:= \left\{ \begin{array}{ll}
       \! \nu(B_\frac{\eps}{2}(x_i)) & (\nu(B_{\frac{\eps}{4}}(x_i)) \geq 0)
        \\
       \! 0 & ({\rm otherwise})
      \end{array} \right. \nts{5} , 
  \end{equation}
  i.e.~the $\frac{\eps}{2}$-intervals of $\mu$ and $\nu$ are enlarged by
  $\frac{\eps}{4}$ corresponding to $\kappa = \frac{1}{2}$. This choice
  facilitates the proof and has no influence on the resulting $D_q$ as Riedi
  has shown (cf \cite{Riedi}). Let $i\in\Z$ with $(\mu \ast \nu)_i > 0$,
  i.e.~the integral over the $i$-th interior interval is non-zero. When
  calculating $(\mu \ast \nu)_i$ we integrate over the wide diagonal strip in
  figure \ref{fig2}(b). The large squares $B_{\frac{\eps}{2}}(x_{2i+2j})
  \times B_{\frac{\eps}{2}}(x_{2i-2j})$, $j \in \Z$, are disjoint and are all
  contained in the strip. Therefore,
  \begin{equation}
    (\mu\ast\nu)_i \geq \sum_j \mu_{2i+2j}\nu_{2i-2j} .
  \end{equation}
  This applies analogously to the other shown five squares and their by
  $(n\eps, -n\eps)$ diagonally translated disjoint copies such that
  \begin{equation}
    6 (\mu \ast \nu)_i \geq \sum_j \sum_{k=-1}^{1} \mu_{2i+j+k} \nu_{2i-j} .
  \end{equation}
  The integral over the narrow diagonal strip determines that $(\mu \ast
  \nu)_i$ is greater than zero. In the same way the integral over the small
  squares determines whether the terms on the right hand side are greater than
  zero. As the narrow strip is contained in the union of the small squares,
  the right hand side is greater than zero as $(\mu \ast \nu)_i$ is.
  We therefore can take the $q$-th power on both sides and (omitting all
  terms being zero) use $(\sum_i
  x_i)^q \leq \sum_i x_i^q$ for $q<1$ and arbitrary positive numbers $x_i$
  to obtain
  \begin{equation}
    6^q (\mu\ast\nu)_i^q \leq \sum_j \sum_{k=-1}^{1} \mu_{2i+j+k}^q
    \nu_{2i-j}^q .
  \end{equation}
  When summing over all $i$ with $(\mu\ast\nu)_i > 0$, each combination
  $\mu_i^q \nu_j^q$ appears at most twice. Furthermore, adding terms which do
  not already appear only enlarges the right hand side. Therefore,
  \begin{equation}
    6^q \sum_i (\mu\ast \nu)_i^q \leq 2 \sum_i \mu_i^q \sum_j \nu_j^q .
  \end{equation}
  From this we immediately obtain
  \begin{equation}
    \frac{1}{q-1} \frac{\ln \sum_i (\mu\ast \nu)_i^q + \ln 6^q}{\ln \eps}
    \leq \frac{1}{q-1} \frac{\ln \sum_i \mu_i^q + \ln \sum_j \nu_j^q +\ln
    2} {\ln \frac{\eps}{2} +\ln 2}
  \end{equation}
  which implies $D_q(\mu \ast \nu) \leq D_q(\mu)+D_q(\nu)$ in the limit $\eps
  \to 0$.
\end{proof}
Note that this proof easily generalizes to measures on $\R^n$.
In \cite{Riedi} the invariance of the $D_q$-spectrum with respect to
bi-Lipschitz maps was shown, i.e.~if $f: \R \to \R$ is a bi-Lipschitz map then
\begin{equation}
  D_q(\mu) = D_q(f_\#(\mu)). \label{eqn7}
\end{equation}
As in the case of the pointwise dimension it is sufficient that the function
$f$ is bi-Lipschitz on an interval containing the support of $\mu$.  Therefore,
we can immediately deduce from theorem \ref{theorem1} and (\ref{eqn7}) that
\begin{eqnarray}
    D_q(\mu^{(m)}) &= D_q(\mu^{(x)} \ast A_\# \mu^{(x)}) \nonumber \\
    &\leq D_q(\mu^{(x)}) + D_q(A_\# \mu^{(x)}) = 2 D_q(\mu^{(x)}) .
\end{eqnarray}
As the $D_q$-spectrum of the invariant measure of the effective field is --
at least on a numerical level -- very well known (cf
\cite{Nowotny}) this provides interesting insights for the
$D_q$-spectrum of the measure of the local magnetization (cf figure
\ref{fig7}). In section \ref{numericsec} we will discuss how to obtain the
$D_q$-spectrum of the measure of the local magnetization numerically and we will
compare the results to the bounds obtained in this section.

\section{Convolution of measures on Cantor sets} \label{cantorsec}
\begin{figure}
\psfrag{-0.6}{\raisebox{-1mm}{$\!\!\sst -0.6$}}
\psfrag{0.6}{\raisebox{-1mm}{$\!\!\sst 0.6$}}
\psfrag{0}{\raisebox{-1mm}{$\!\sst \!0$}}
\psfrag{-0.8}{\raisebox{-1mm}{$\!\!\sst -0.8$}}
\psfrag{0.8}{\raisebox{-1mm}{$\!\!\sst 0.8$}}
\psfrag{8}{\raisebox{-1mm}{$\!\sst \!8$}}
\psfrag{3}{\raisebox{-1mm}{$\!\sst \!3$}}
\psfrag{30}{\raisebox{-1mm}{$\!\!\sst \!30$}}
\psfrag{10}{\raisebox{-1mm}{$\!\!\sst \!10$}}
\parbox[c]{0.305\textwidth}{\epsfig{file=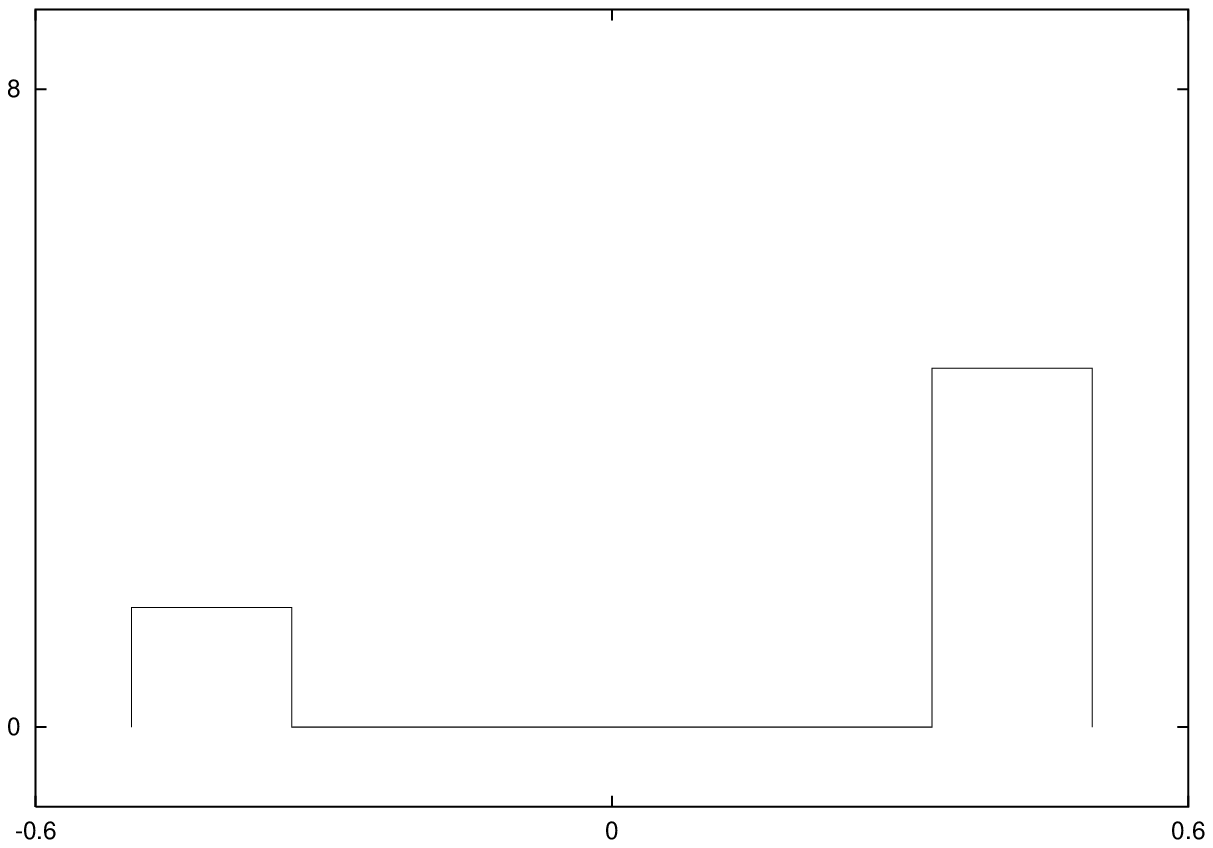, width= 0.3\textwidth}}
$\ast$
\parbox[c]{0.305\textwidth}{  \epsfig{file=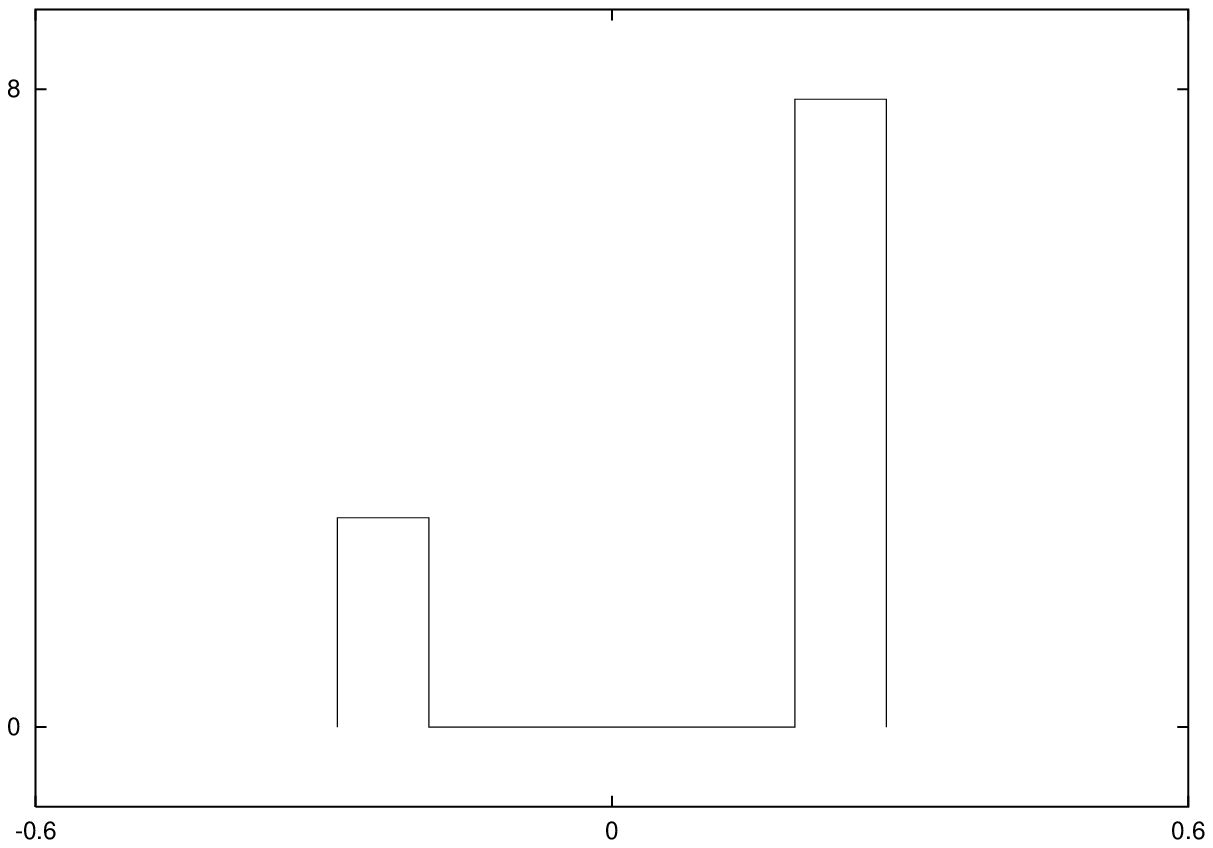, width= 0.3\textwidth}}
$=$
\parbox[c]{0.305\textwidth}{
  \epsfig{file=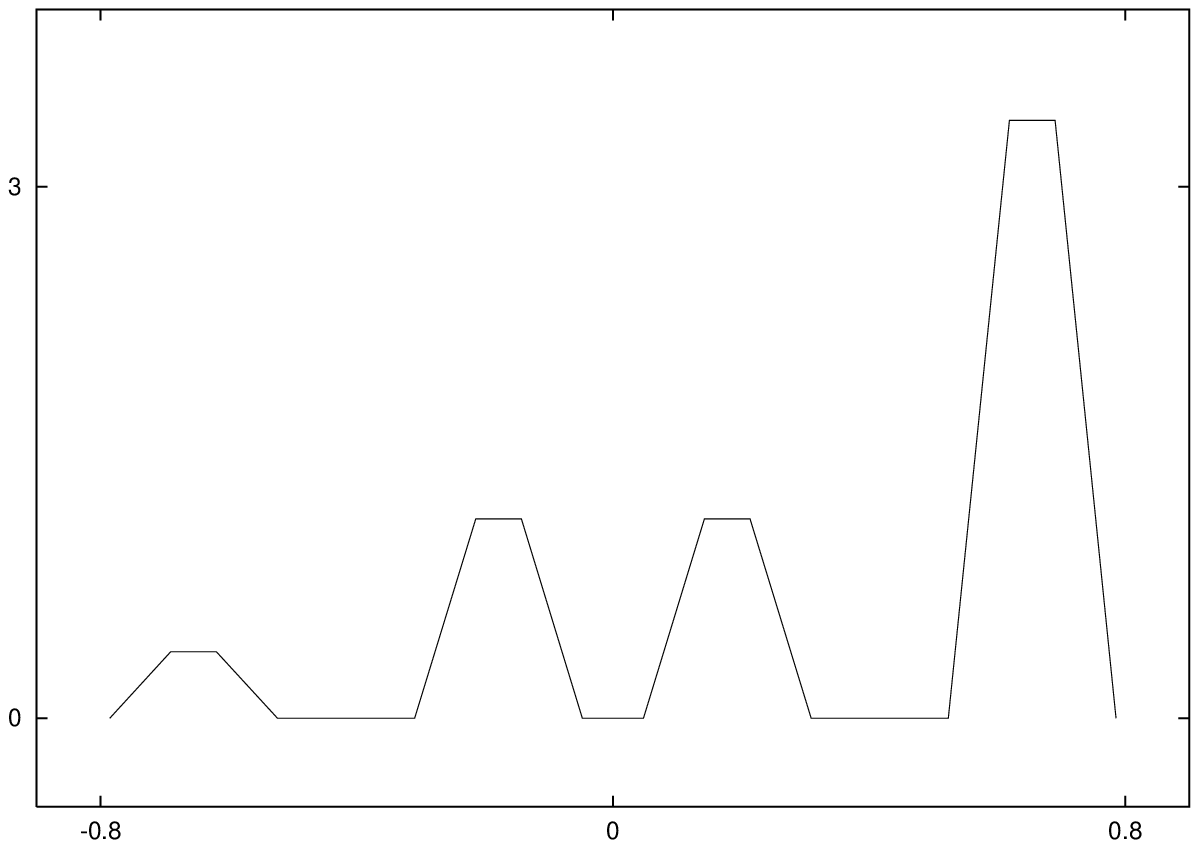, width= 0.3\textwidth}} \\[0.2cm]
\parbox[c]{0.305\textwidth}{\epsfig{file=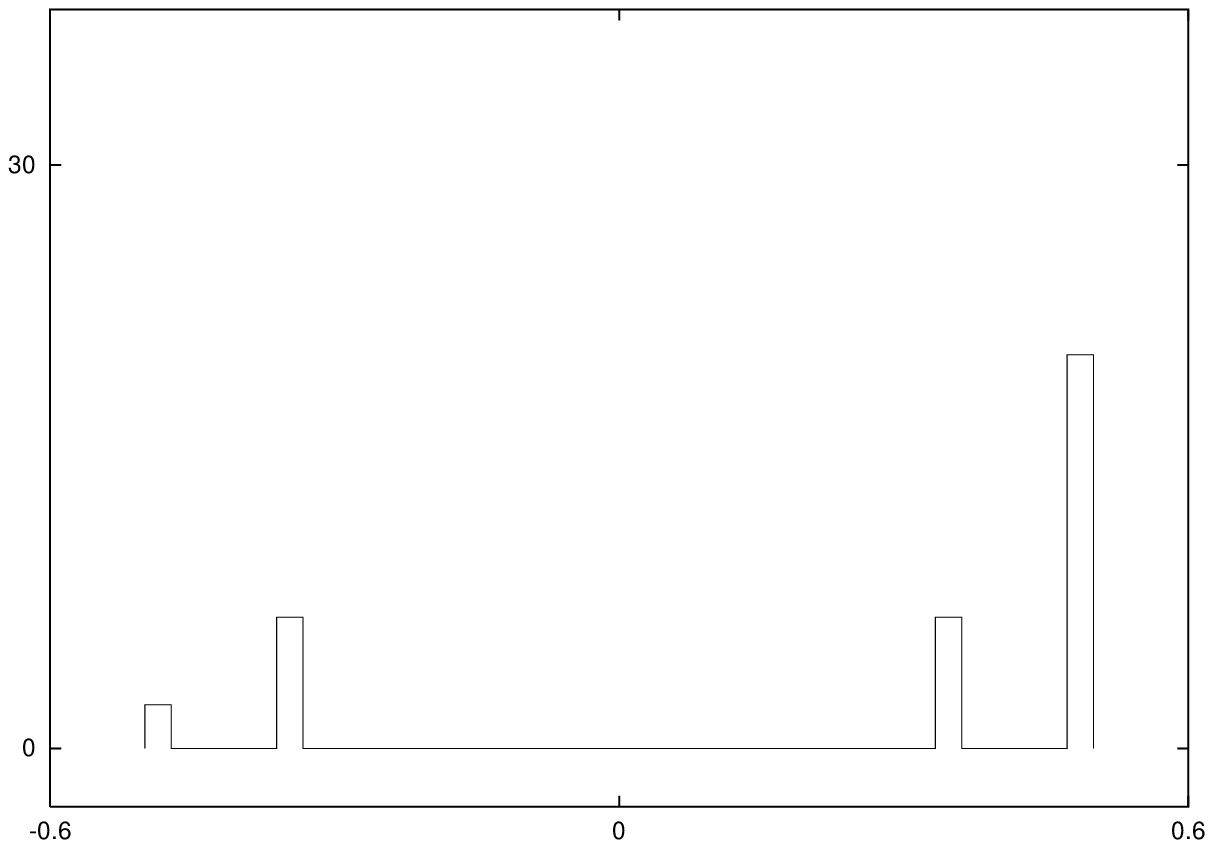, width= 0.3\textwidth}}
$\ast$ 
\parbox[c]{0.305\textwidth}{\epsfig{file=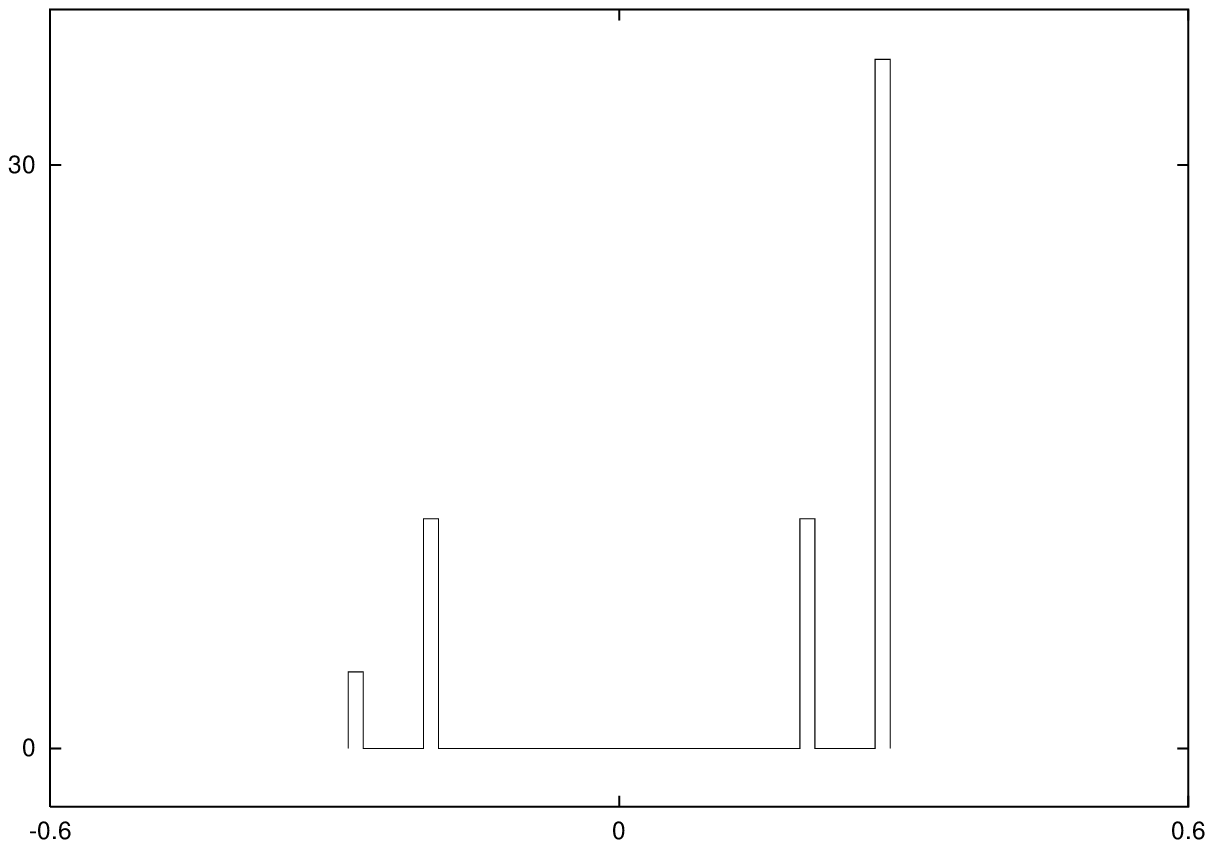, width= 0.3\textwidth}}
$=$
\parbox[c]{0.305\textwidth}{\epsfig{file=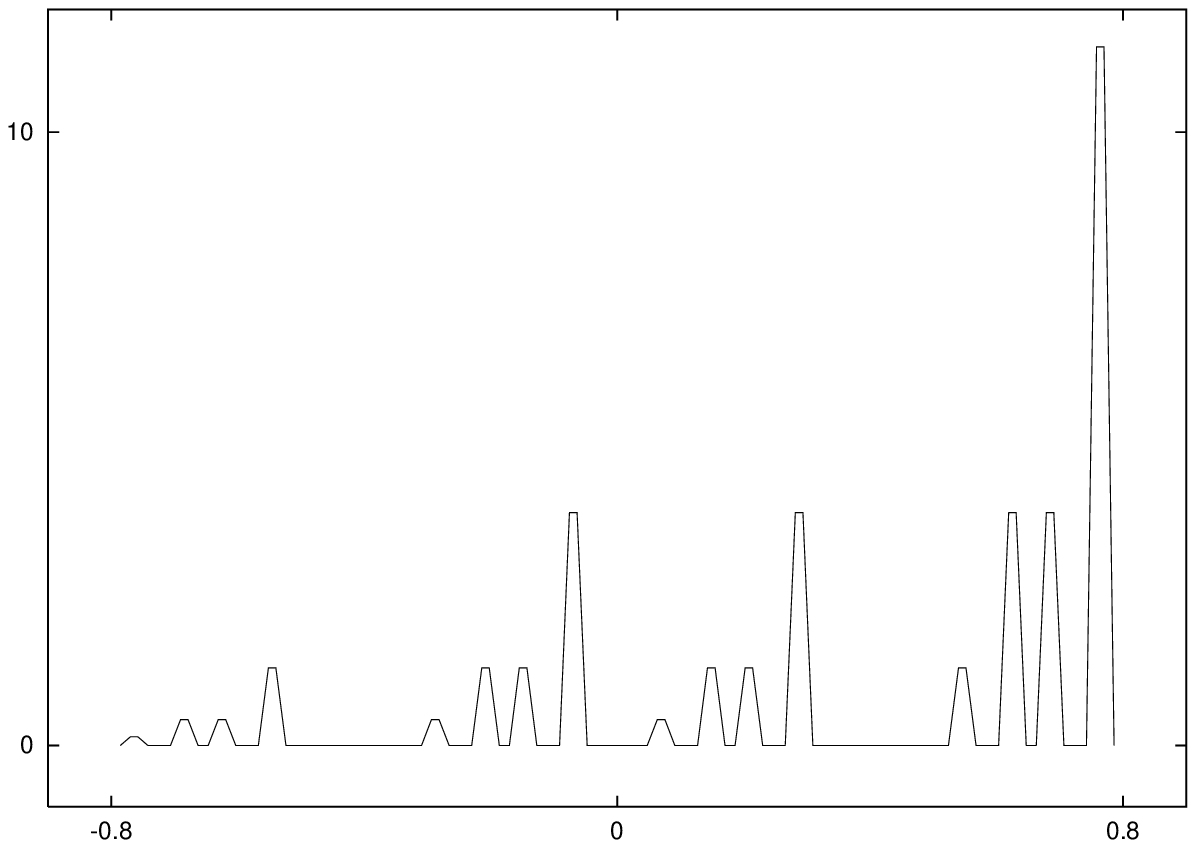, width= 0.3\textwidth}}
\caption{Illustration of the convolution of $\msu{1}{a,p}$ with $\lambda_\#
  \msu{1}{a,p}$ and $\msu{2}{a,p}$ with $\lambda_\# \msu{2}{a,p}$ for $a=
  \frac{1}{6}$, $p= \frac{3}{4}$ and $\lambda= \frac{4}{7}$. The condition
  for the disjointness of the trapezoids, $a/(1-2a) \leq \lambda \leq (1-2a)$
  is clearly fulfilled for this choice of parameters such that in the limit
  $n \to \infty$ example \ref{example1} applies. \label{fig3}}
\end{figure}

\begin{figure}
\psfrag{-0.6}{\raisebox{-1mm}{$\!\!\sst -0.6$}}
\psfrag{0.6}{\raisebox{-1mm}{$\!\!\sst 0.6$}}
\psfrag{0}{\raisebox{-1mm}{$\!\sst \!0$}}
\psfrag{-1}{\raisebox{-1mm}{$\!\!\sst -1$}}
\psfrag{1}{\raisebox{-1mm}{$\!\!\sst 1$}}
\psfrag{1.5}{\raisebox{-1mm}{$\!\!\!\sst \!1.5$}}
\psfrag{2.5}{\raisebox{-1mm}{$\!\!\!\sst \!2.5$}}
\psfrag{2}{\raisebox{-1mm}{$\!\sst \!2$}}
\psfrag{5}{\raisebox{-1mm}{$\!\sst \!5$}}
\parbox[c]{0.305\textwidth}{\epsfig{file=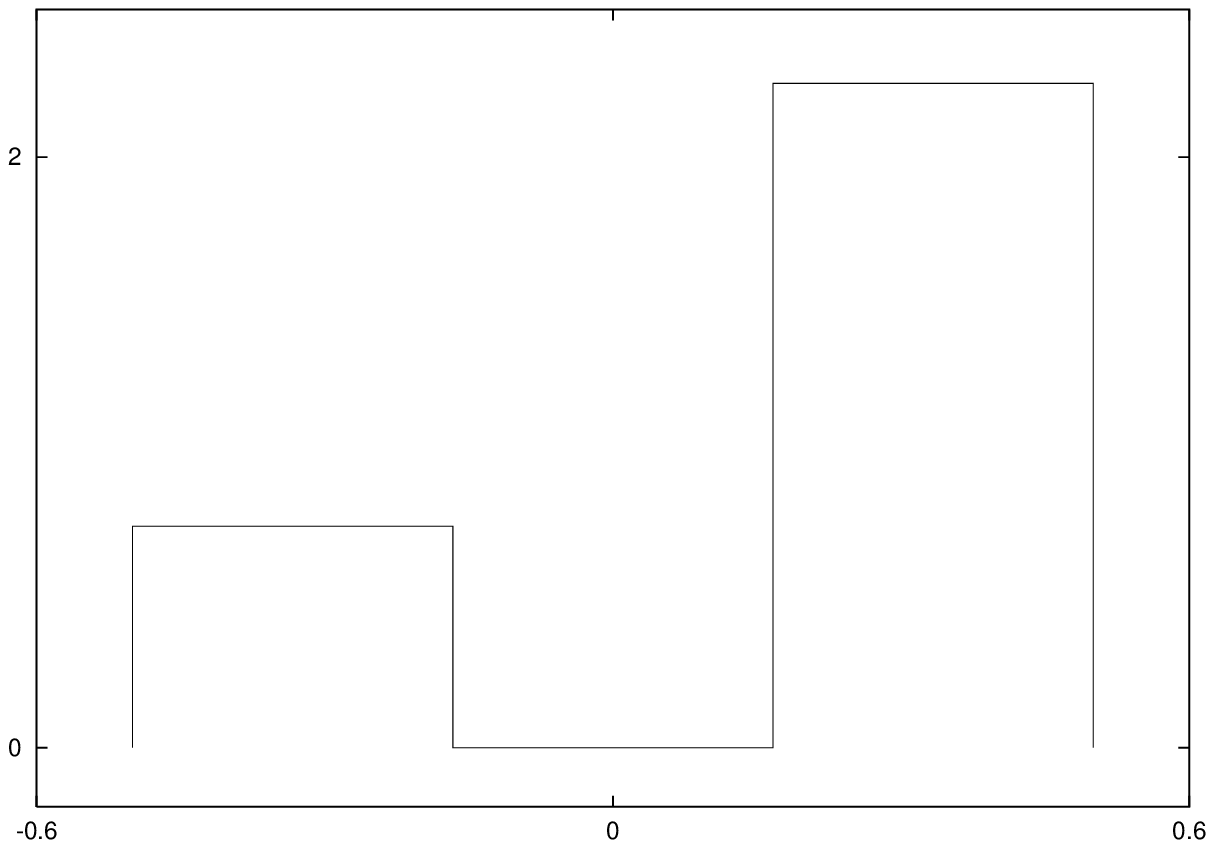, width= 0.3\textwidth}}
$\ast$
\parbox[c]{0.305\textwidth}{  \epsfig{file=fig4a.eps, width= 0.3\textwidth}}
$=$
\parbox[c]{0.305\textwidth}{\epsfig{file=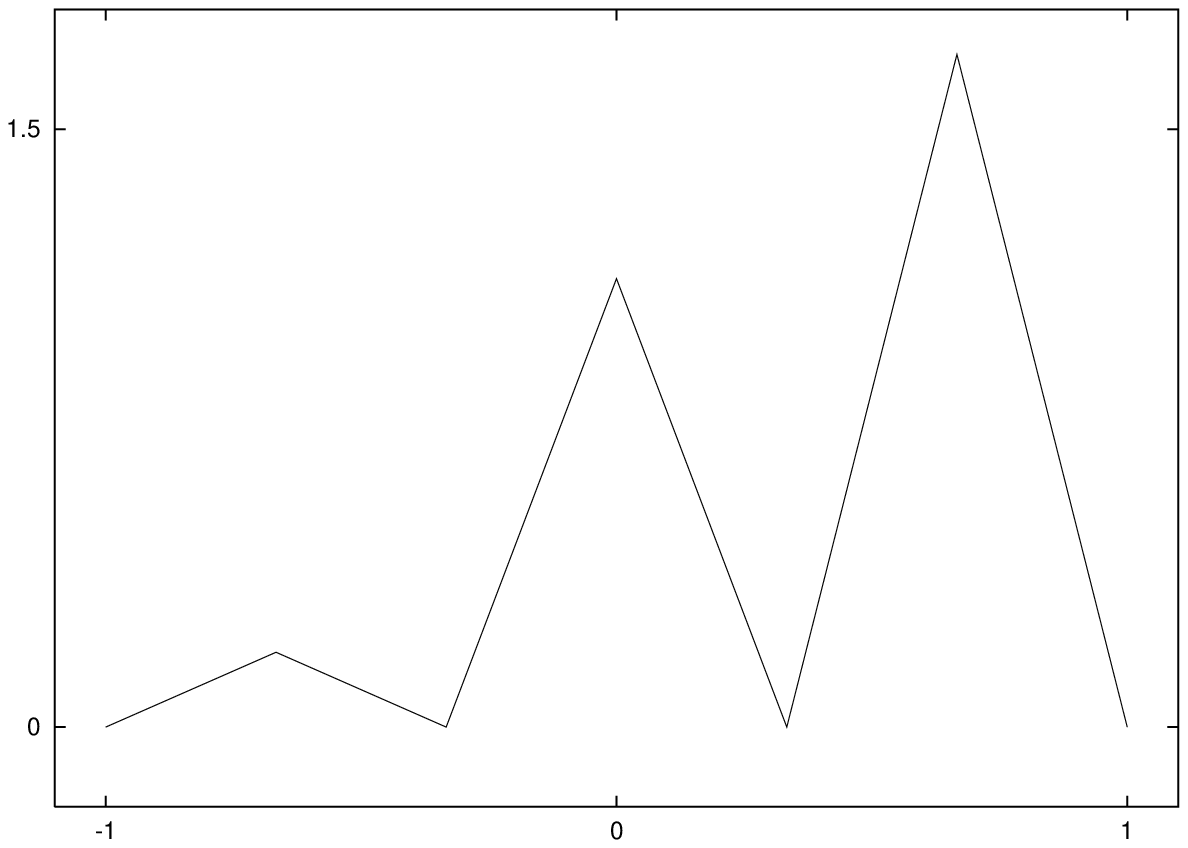, width= 0.3\textwidth}}
\\[0.2cm]
\parbox[c]{0.305\textwidth}{\epsfig{file=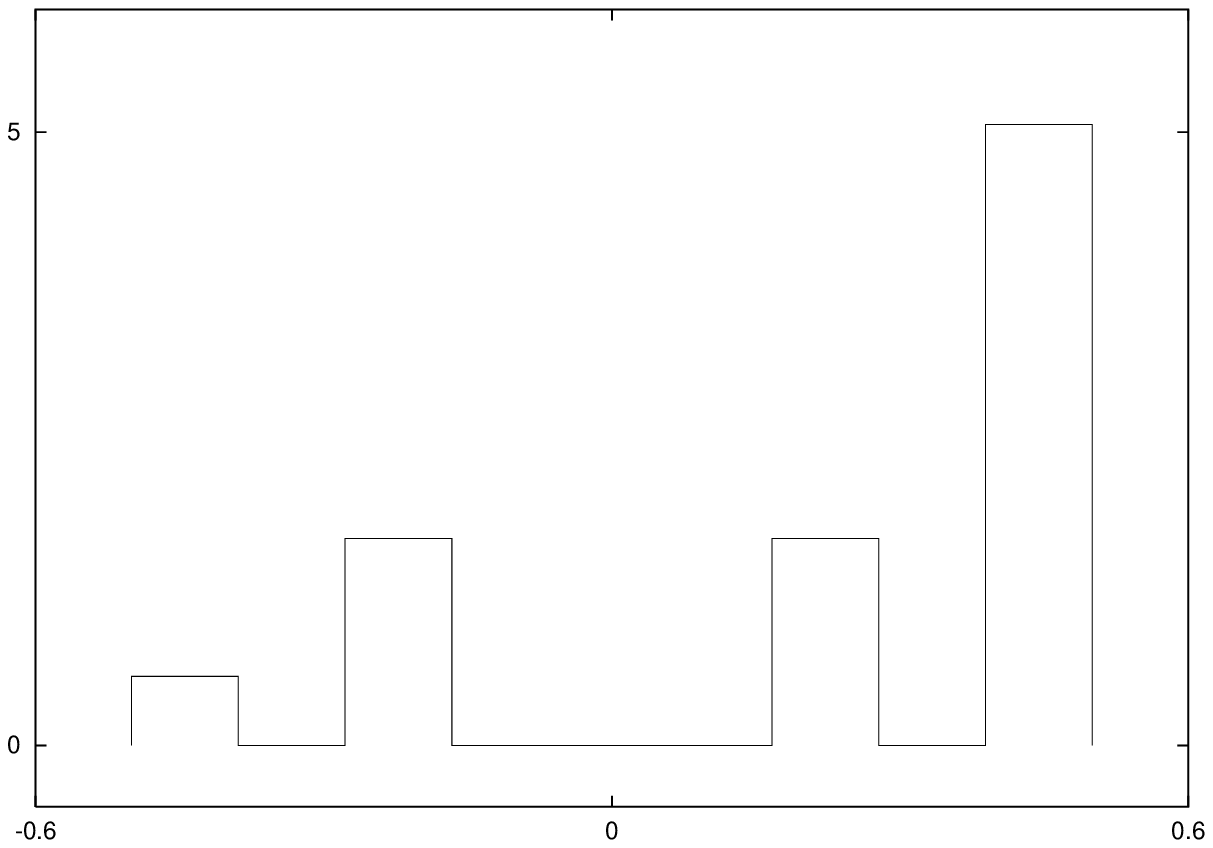, width= 0.3\textwidth}}
$\ast$ 
\parbox[c]{0.305\textwidth}{\epsfig{file=fig4d.eps, width= 0.3\textwidth}}
$=$
\parbox[c]{0.305\textwidth}{\epsfig{file=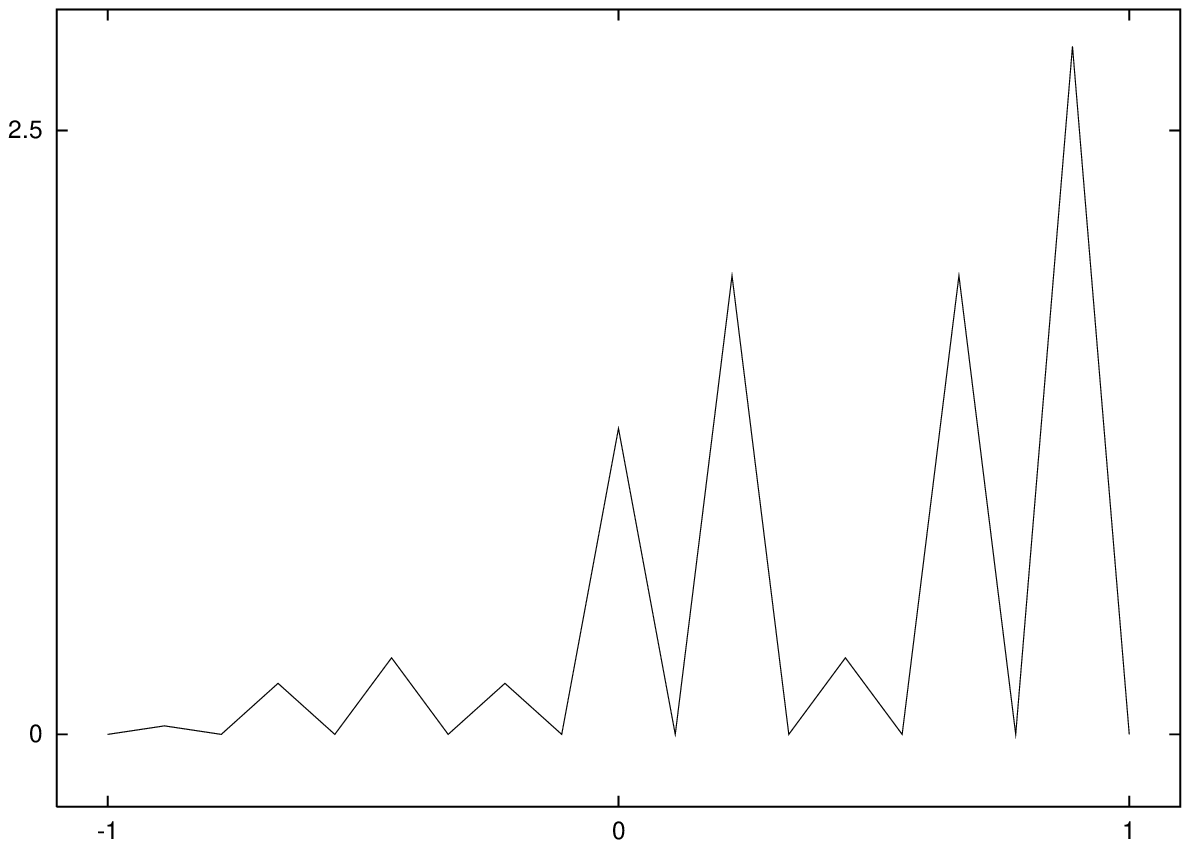, width= 0.3\textwidth}}
\caption{Illustration of the convolution of $\msu{1}{a,p}$ with itself and
  $\msu{2}{a,p}$ with itself for $a= \frac{1}{3}$ and $p= \frac{3}{4}$. For
  this choice of parameters example \ref{example2} applies in the limit $n
  \to \infty$. \label{fig4}}

\end{figure}

\begin{figure}
\psfrag{0.5}{\fnt $\;\;\, 0.5$}
\psfrag{1}{\fnt $1$}
\psfrag{1.5}{\fnt $\;\;\, 1.5$}
\psfrag{-8}{\fnt $-8$}
\psfrag{-6}{\fnt $-6$}
\psfrag{-4}{\fnt $-4$}
\psfrag{-2}{\fnt $-2$}
\psfrag{l}{$\ln \lambda$}
\psfrag{Dq}{\raisebox{0.4cm}{\rotatebox{180}{$D_q$}}}
\epsfig{file=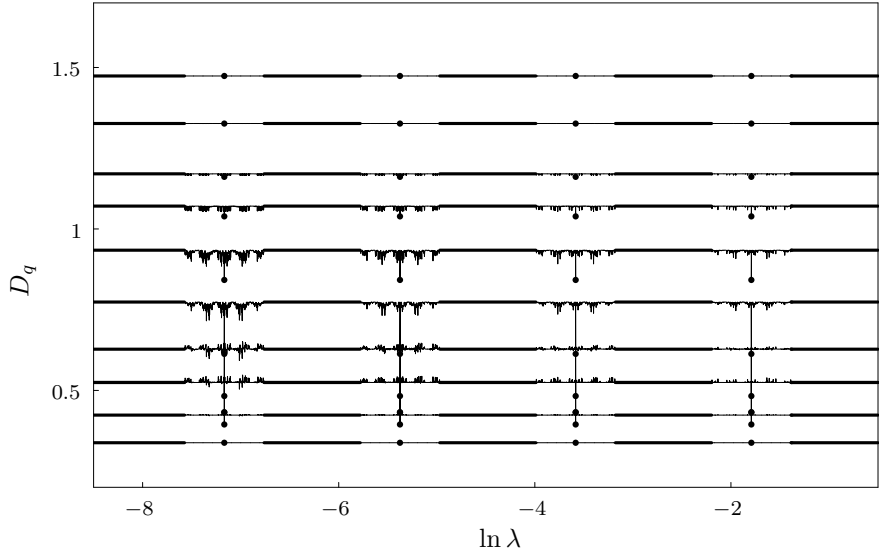, width=0.9\textwidth}
\caption{$D_q$ spectrum of $\mu_{a,p} \ast \lambda_\# \mu_{a,p}$ as a
  function of $\ln \lambda$ for $q= -20$, $-6$, $-3$, $-2$, $-1$,
  $0$, $1$, $2$, $4$ and $20$. The thick lines are the exact $D_q$ obtained in
  example \ref{example1} and the points the $D_q$ obtained in example
  \ref{example2}. The thin lines are numerical results obtained from
  iteration depths $l=r=8$ compared to $l=r=7$ in the new natural
  partition method (cf section \ref{dqmethods}). \label{fig5}}
\end{figure}

Let $\cantor{0}{a}= [-\frac{1}{2}, \frac{1}{2}]$ and $\cantor{n}{a}$ be
defined inductively by $\cantor{n}{a} := f_{a+} (\cantor{n-1}{a}) \cup
f_{a-}(\cantor{n-1}{a})$ with $f_{a+}(x) = ax+\frac{1-a}{2}$ and $f_{a-}= ax
- \frac{1-a}{2}$. The infinite intersection $\limcan{a} :=
\bigcap_{n=0}^{\infty} \cantor{n}{a}$ is the $a$-Cantor set. On the
approximating sets $\cantor{n}{a}$ we
define the probability densities
\begin{equation}
  \prob{n}{a,p}(x) = \frac{p}{a} \prob{n-1}{a,p} (f_{a+}^{-1}(x))
  + \frac{1-p}{a} \prob{n-1}{a,p} (f_{a-}^{-1}(x)) .
\end{equation}
The corresponding measures are denoted by $\msu{n}{a,p}(X) = \int_X
\prob{n}{a,p} dx$ for any $X \in {\mathcal B}(\R)$. The measures
$\msu{n}{a,p}$ converge to a limit measure $\mu_{a,p}$ which is often
referred to as an $a$-Cantor set with weights $p$ and $1-p$. For a generic
choice of $a$ and $p$ the measure $\mu_{a,p}$ is a multifractal. For an
illustration cf figure \ref{fig3} and \ref{fig4}. In the following example we
calculate the $D_q$ spectrum of the convolution of $\mu_{a,p}$ with
$\lambda_\# \mu_{a,p}$, a ``compressed'' version of itself ($\lambda \leq
1$).  As this is in general a hard problem we discuss two examples.
%
%%% Local Variables: 
%%% mode: latex
%%% TeX-master: "article"
%%% End: 
\begin{example} \label{example1}
  Let $0 < p < 1$ and $a/(1-2a) \leq \lambda \leq (1-2a)$. This is meaningful
  for $a \leq \frac{1}{4}$. Then, the two intervals of $\supp \msu{1}{a,p}$
  fit into the gap of $\supp \lambda_\# \msu{1}{a,p}$ and on the other hand
  the complete $\supp \lambda_\# \msu{1}{a,p}$ fits into the gap of $ \supp
  \msu{1}{a,p}$. Self-similarity of $\limcan{a}$ and $\lambda_\# \limcan{a}$
  imply that for any given $n$ and $y \in \R$ at most one pair of bars in
  $\prob{n}{a,p}(x)$ and $\lambda_\# \prob{n}{a,p}(y-x)$ can overlap. The
  convolution is therefore a collection of trapezoids as shown in figure
  \ref{fig3}. We denote the intervals of $\supp \msu{n}{a,p} \ast \lambda_\#
  \msu{n}{a,p}\,$ (the bases of the trapezoids) by $\trap{n}{j}$, $j= 1,
  \ldots, 2^n$. For larger
  $n$ only the structure of the measure within the trapezoids but not their
  total measure changes, i.e.~$\mu_{a,p} \ast \lambda_\# \mu_{a,p}
  (\trap{n}{j}) = \msu{n}{a,p} \ast \lambda_\# \msu{n}{a,p} (\trap{n}{j})$.
  We therefore can use $\mu_{a,p} \ast \lambda_\# \mu_{a,p}$ and
  $\msu{n}{a,p} \ast \lambda_\# \msu{n}{a,p}$ interchangeably. This fortunate
  circumstance is due to the fact that the self-similarity of the Cantor sets
  induces a direct iteration for the convolution making it self-similar itself.
  The analytical treatment of the $D_q$-spectrum in this example is essentially 
  based on this fact. If we choose
  $\eps_n:= a^n + \lambda a^n$, which is the width of the trapezoids at level
  $n$, boxes $B_{\frac{3}{2}\eps_n}(x_i)$, $x_i = i \eps_n$, $i \in \Z$ with
  $\mu_{a,p}(B_{\frac{\eps_n}{2}}(x_i)) > 0$ contain at least one whole
  trapezoid of level $n$ and intersect at most four.  Thus, denoting
  \begin{equation}\fl
    \mu_i := \left\{\begin{array}{ll} \mu_{a,p} \ast \lambda_\# \mu_{a,p}
        (B_{\frac{3}{2}\eps_n}(x_i))
        & (\mu_{a,p} \ast \lambda_\# \mu_{a,p} (B_{\frac{\eps_n}{2}}(x_i)) > 0) \\
        0 & (\rm otherwise) \end{array} \right.
  \end{equation}
  we obtain for $q > 0$, $q \not= 1$
  \begin{equation}
    \mu_{a,p} \ast \lambda_\# \mu_{a,p} \big(\trap{n}{j(i)}\big)^q \leq
    \mu_i^q \leq \big(4 \max_{j \in J(i)}
    \mu_{a,p} \ast \lambda_\# \mu_{a,p} (\trap{n}{j})\big)^q  ,
  \end{equation}
  where $j(i)$ is the index of a trapezoid completely contained in
  $B_{\frac{3}{2} \eps_n}(x_i)$ and $J(i)$ is the set of the indices of all
  trapezoids intersecting $B_{\frac{3}{2} \eps_n}(x_i)$.
  As any trapezoid can appear at most four times on the right hand side
  when summing over $i$ and any trapezoid appears at least once on the
  left hand side this implies
  \begin{equation}
    \fl
    \sum_j \mu_{a,p} \ast \lambda_\# \mu_{a,p} \big(\trap{n}{j}\big)^q
    \leq \sum_i \mu_i^q
    \leq 4\sum_j \big(4 \mu_{a,p} \ast \lambda_\# \mu_{a,p} (\trap{n}{j})\big)^q
  \end{equation}
  The measures of the trapezoids can explicitly be calculated such that
  \begin{equation}
    \fl
    \sum_j \mu_{a,p} \ast \lambda_\# \mu_{a,p} (\trap{n}{j})^q
    = \sum_{k,l = 0}^n \binom{n}{k}
    \binom{n}{l} 
    (p^k(1-p)^{n-k} p^l (1-p)^{n-l} )^q .
  \end{equation}
  Applying $\sum_k \binom{n}{k} (p^q)^k((1-p)^q)^{n-k}= (p^q + (1-p)^q)^n$
  we obtain
  \begin{equation}
    (p^q+(1-p)^q)^{2n} \leq \sum_i \mu_i^q \leq 4^{q+1}
    (p^q+(1-p)^q)^{2n}
  \end{equation}
  and therefore
  \begin{equation}
    D_q(\mu_{a,p})= \frac{1}{q-1} \frac{2\ln(p^q+(1-p)^q)}{\ln a}
    . \label{eqn3} 
  \end{equation}
  For $q < 0$ the argument is the same with reversed inequality signs
  which leads to the same result. \\
  For $q=1$ we calculate the limit $q \to 1$ of (\ref{eqn3}) yielding
  \begin{eqnarray}
    D_1 &= \lim_{q \to 1} \frac{1}{q-1} \frac{2\ln(p^q+(1-p)^q)}{\ln a}
    \nonumber \\
    &= 2 (p\ln p + (1-p)\ln (1-p))/ \ln a . \label{eqn3b} 
  \end{eqnarray}
  For any $\lambda \in [a^{k+1}/(1-2a), a^{k}(1-2a)]$, $k \in \N$, the
  arguments above apply to $\msu{n+k}{a,p} \ast \lambda_\#\msu{n}{a,p}$ which
  according to lemma \ref{lemma1} also converges to $\mu_{a,p} \ast
  \lambda_\# \mu_{a,p}$. Therefore,
  (\ref{eqn3}) and (\ref{eqn3b}) apply to all $\lambda$ taken from these
  intervals. For an example cf figure \ref{fig5}.
\end{example}

%%% Local Variables: 
%%% mode: latex
%%% TeX-master: "article"
%%% End: 

\begin{figure}
\psfrag{0}{\fnt $0$}
\psfrag{0.5}{\fnt $0.5$}
\psfrag{1}{\fnt $\! 1$}
\psfrag{1.5}{\fnt $1.5$}
\psfrag{2}{\fnt $\! 2$}
\psfrag{2.5}{\fnt $2.5$}
\psfrag{3}{\fnt $\! 3$}
\psfrag{4}{\fnt $\! 4$}
\psfrag{Dq}{\raisebox{0.3cm}{$D_q$}}
\psfrag{lna}{$-\ln a$}
\centerline{\epsfig{file=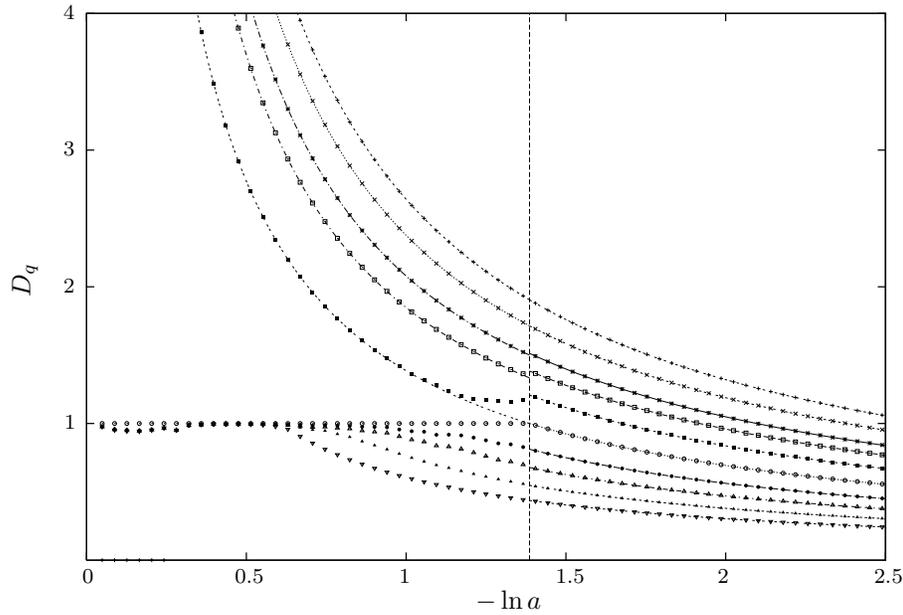, width= 0.9\textwidth}}
\caption{$D_q$-spectrum of $\mu_{a,p} \ast \lambda_\# \mu_{a,p}$ as a
  function of $-\ln a$ with $p= \frac{3}{4}$, $\lambda= \frac{1}{2}$ and $q=
  -20$, $-6$, $-3$, $-2$, $-1$, $0$, $1$, $2$, $4$, $20$. The
  lines to the left of $-\ln a= \ln 4$ are the usual lower bounds based on the
  pointwise dimension at the left boundary of $\limcan{a,p}$. The lines on the
  right are the calculated exact values of $D_q$ according to (\ref{eqn3})
  and (\ref{eqn3b}). The points are numerical results for iteration depths
  $l=r=6$ compared to $l=r=5$. \label{fig6}}
\end{figure}

\begin{example} \label{example2}
  Let $0 < p < 1$, $a \leq \frac{1}{3}$ and $\lambda= 1$. We then
  choose $\eps_n:= 2a^n$ and boxes of length $\eps_n$ in such a way that each
  box covers one of the spikes of $\msu{n}{a,p} \ast \msu{n}{a,p}$. This is a
  permitted choice and avoids any complications with $q < 0$ such that we
  refrain from using enlarged boxes here.  The situation is like
  shown in figure \ref{fig4}, i.e.~for $n=1$, $(\msu{1}{a,p} \ast
  \msu{1}{a,p})_i= (1-p)^2$, $2p(1-p)$ and $p^2$ for $i= 1$, $2$ and $3$
  respectively. As in example \ref{example1} self-similarity implies for our
  specific choice of boxes that we can use $\msu{n}{a,p} \ast \msu{n}{a,p}$
  and $\mu_{a,p} \ast \mu_{a,p}$ interchangeably as they coincide on all
  boxes. Furthermore, the result of the convolution of the next iteration,
  $n=2$, can be constructed by replacing each triangle in figure
  \ref{fig4} by the complete figure and choosing the corresponding weight.
  Therefore, the $(\msu{2}{a,p} \ast \msu{2}{a,p})_i$, $i=1, \ldots, 9$, sum
  up to
  \begin{eqnarray}
    & \sum_{k=0}^2 \sum_{l=0}^{2-k}
    \binom{2}{k} \binom{2-k}{l} (1-p)^{2k}(2p(1-p))^{2-k}
    p^{2(2-k-l)}  \\
    &= \left((1-p)^2 + 2p(1-p)+ p^2\right)^2 .
  \end{eqnarray}
  This continues for larger $n$ such that we have
  \begin{eqnarray}
    \fl
    \sum_i (\msu{n}{a,p} \ast \msu{n}{a,p})_i^q &= \sum_{k=0}^n \sum_{l=
    0}^{n-k} \binom{n}{k} \binom{n-k}{l} \left( (1-p)^{2k} (2p(1-p))^{n-k}
    p^{2(n-k-l)}\right)^q \\
  &= ((1-p)^{2q} + 2^qp^q(1-p)^q + p^{2q})^n .
  \end{eqnarray}
  For $q \not= 1$ this yields
  \begin{eqnarray}
    D_q &= \frac{1}{q-1} \lim_{n \to \infty} \frac{n \ln( (1-p)^{2q} +
    2^qp^q(1-p)^q + p^{2q})}{n \ln a +\ln2} \\
  &= \frac{1}{q-1} \frac{\ln( (1-p)^{2q} +
    2^qp^q(1-p)^q + p^{2q})}{\ln a}. \label{eqn5}
  \end{eqnarray}
  The limit $q\to 1$ results in
  \begin{eqnarray}
    D_1 &=\lim_{q\to 1} \frac{1}{q-1} \frac{\ln( (1-p)^{2q} +
    2^qp^q(1-p)^q + p^{2q})}{\ln a} \nonumber \\
  &= 2 (p\ln p + p(1-p)\ln 2+
    (1-p) \ln(1-p))/\ln a . \label{eqn5b}
  \end{eqnarray}  
  As in example \ref{example1} the arguments can be repeated for the
  $a^n$-fold value of $\lambda$ such that (\ref{eqn5}) and (\ref{eqn5b}) are
  also correct for all $\lambda= a^n$, $n \in \N$. This means that we can
  calculate the $D_q$-spectrum at specific points between the intervals where
  it is given by (\ref{eqn3}) and (\ref{eqn3b}).
  The results are illustrated for $a= \frac{1}{6}$ and $p= \frac{3}{4}$ in
  figure \ref{fig5}. 
\end{example}
In the case of the random-field Ising model we are interested in the
dependence of the $D_q$-spectrum on the strength $h$ of the random field
which rather corresponds to varying $a$ in the convolution of Cantor sets.
Viewing (\ref{eqn3}) and (\ref{eqn3b}) in this light we can choose $\lambda=
\frac{1}{2}$ such that for all $a < \frac{1}{4}$ example \ref{example1}
applies. This results in the right part of the $D_q$-spectrum shown in figure
\ref{fig6}. For the left part with $a$ close to one we have the usual lower
bounds for $q < 0$ based on the pointwise dimension at the boundary of the
support and $D_q = 1$ for $q \geq 1$ from the regularity of $\mu_{a,p}$.

%%% Local Variables: 
%%% mode: latex
%%% TeX-master: "article"
%%% End: 

%%% Local Variables: 
%%% mode: latex
%%% TeX-master: "article"
%%% End: 

The convolution of two-scale Cantor sets shows a far richer behaviour than the two
examples above. It turns out that the lacking strict self-similarity does not
allow the kind of analytical treatment we used so far. Numerical
investigations show that the $D_q$-spectrum strongly depends on the two
scales of the Cantor sets. Furthermore, in a large parameter regime the
obtained numerical estimates are extremely sensitive to the iteration depth
to which the Cantor sets are generated. The situation is of similar
complexity as for the measure of the local magnetization in the
one-dimensional random-field Ising model discussed below. We therefore
refrain from discussing this case and turn to the characterization of the
measure of the local magnetization.

\section{$\mathbf D_{\mathbf q}$-spectrum of the local magnetization} \label{numericsec}
In this section we present our numerical methods and the resulting
$D_q$-spectra for the measure of the local magnetization in the random-field
Ising model. We obtain the $D_q$-spectrum of the measure of the local
magnetization in several steps.
\subsection{Generation of the measure of the effective field}
The measure of the local magnetization in the bulk of the random-field Ising
chain is given by (\ref{magnet}), i.e.~it is the convolution of the measure
of the effective field $x_n$ with a distorted version of itself. Therefore,
as a first step for a numerical treatment, an approximation of this measure
is needed. It turns out that a very useful approximation is the following (cf
\cite{Nowotny, Behn6}). We take the partition of the invariant interval $I=
[x^*_-, x^*_+]$ given by the points $\{x_i\}= \{ f_{\{\sigma\}_n} (x^*_-),
f_{\{\sigma\}_n} (x^*_+)\}$. It is sometimes called the ``new natural
partition'' (cf \cite{Behn6}). The points $x^*_-$ and $x^*_+$ are the fixed
points of the iteration (\ref{mapping}). The symbols $f_{\{\sigma\}_n}(x)$
denote the iterated functions $f_{\{\sigma\}_n}(x):= f_{\sigma_1} \circ
f_{\sigma_2} \circ \ldots \circ f_{\sigma_n}(x)$ with $f_+(x)= A(x) + h$,
$f_-(x)= A(x) -h$ and $ \{\sigma\}= \{\sigma_1, \sigma_2, \ldots, \sigma_n\}$
with $\sigma_i \in \{+,-\}$.  The $n$-fold application of the
Frobenius-Perron equation (\ref{frobenius}) on some initial distribution
$P_0$ yields \cite{Behn6}
\begin{equation}
  P_n(x)= \frac{1}{2^n} \sum_{\{\sigma\}_n} P_0(f_{\{\sigma\}_n}^{-1}(x)) 
\end{equation}
where the sum is taken over all symbolic sequences $\{\sigma\}_n$ for
which $f_{\{\sigma\}_n}^{-1}(x)$ exists. In the numerical treatment we 
start with an equipartition on $I$,
\begin{equation}
  P_0= \left\{ \begin{array}{ll} 0 & (x < x^*_-) \\
    (x-x^*_-)/|I| & (x \in I) \\ 1 & (x > x^*_+) \end{array} \right. .
\end{equation}
We then calculate the measure $\mu^{(x)}_n([x_i , x_{i+1}]) =
P_n(x_{i+1})- P_n(x_i)$ on the intervals of the new natural partition and
approximate the density $p^{(x)}_n$ as constant within these intervals.
The resulting histogram $\tilde{p}_n^{(x)}$ is used for
the convolution.  
\subsection{Convolution}
The next step toward the $D_q$-spectrum of the measure of the local
magnetization is to calculate the convolution (\ref{mumagnet}). We will need to
calculate the measure of some given intervals, say $[m_i , m_{i+1}]$.
It is given by
\begin{eqnarray}
  \mu_{l,r}^{(m)}([m_i , m_{i+1}]) &= \tanh \beta_\# (\msu{x}{l} \ast A_\#
  \msu{x}{r} ) ([m_i , m_{i+1}]) \nonumber \\
  &=  \nts{8}
  \int\limits_{1/\beta \tanh^{-1}(m_i)}^{1/\beta \tanh^{-1} (m_{i+1})}
  \nts{10} dy \,
  p^{(x)}_r(y)  \int dx \, p^{(x)}_l (x - A(y)) . \label{eqn6}
\end{eqnarray}
For $p^{(x)}_r$ and $p^{(x)}_l$ we substitute our piecewise constant
approximations $\tilde{p}^{(x)}_r$ and $\tilde{p}^{(x)}_l$ and denote the
inner integral as $F(y)$. This function is piecewise of the form $\kappa_i
A(y) + \eta_i$ and thus can be represented by th coefficients $\kappa_i$,
$\eta_i$ and the endpoints $y_i$ of the intervals $[y_i , y_{i+1}]$ on which
the particular $\kappa_i$ and $\eta_i$ are valid. The outer integral in
(\ref{eqn6}) is therefore approximately
\begin{equation}
  \sum_i \int_{y_i}^{y_{i+1}} \tilde{p}_r^{(x)}(y) (\kappa_i A(y) + \eta_i) dy .
\end{equation}
As $\tilde{p}^{(x)}_r$ is piecewise constant this integrals can easily be
calculated provided $\int A(y) dy$ is known. This integral is given by
\begin{equation}
  \fl
  \int^x A(y) dy = \frac{1}{4\beta^2} (\polylog_2(-e^{2\beta(x-J)})-\polylog_2(
  -e^{2\beta(x+J)}))-J x + C 
\end{equation}
in which $\polylog_2$ denotes the second polylogarithmic function\footnote{
  The $2$-nd polylogarithmic function is $\polylog_2(z):=
  \sum_{k=1}^{\infty} \frac{z^k}{k^2} = - \int_0^z \frac{\log(1-t)}{t} \,dt$.
  }.
It turns out however that for the short intervals we need to integrate on the
implementation of the polylogarithmic function $\polylog_2$ for double precision numbers is less precise than a
simple fifth order Taylor expansion of $\int A(y) dy$ around the center of
the intervals $[y_i , y_{i+1}]$. We therefore use the expansion in our
numerical studies.

By the methods described thus far we have gained the ability to obtain the
measure $\msu{m}{l,r}$ of any given interval $[m_i , m_{i+1}]$ and not too
large iteration depths $l$ and $r$. We now use
two different methods to estimate the $D_q$-spectrum of $\mu^{(m)}$ based on
this.

\subsection{Determination of the $D_q$-spectrum} \label{dqmethods}
The first method uses coverings of $\supp \msu{m}{l,r}$ with boxes of equal
size.  We choose boxes of size $\eps_k:= \eps_0 \cdot s^k$, $k= 1,2, \ldots, N$,
for some $s < 1$ and  points $x_i = i
\eps_k$, $i\in \Z$. We then calculate 
\begin{equation}
  Z^{(B)}_k = \sum_{(i)} (\msu{m}{l,r} (B_{\frac{3}{2}\eps_k}(x_i)))^q
\end{equation}
in which only indices $i$ fulfilling $\msu{m}{l,r} (B_{\frac{\eps_k}{2}}(x_i)
> 0$ are considered. A linear fit of $\ln Z^{(B)}_k$ as a function of $\ln
\eps_k$ yields $\ln Z^{(B)}_k \sim \tau_q \ln \eps_k + {\rm const.}$ providing
the desired estimate $D_q= \tau_q/(q-1)$ of the generalized fractal dimension
$D_q$.

The second method is based on the stationarity of a suitably chosen partition
function. When observing the process of the convolution of $\msu{x}{l}$ and
$A_\# \msu{x}{r}$ more closely it becomes clear that there is a qualitative
change whenever bands of $\msu{x}{l}$ and bands of $A_\# \msu{x}{r}$ start or
cease to overlap, i.e.~for values $m_i$ of the magnetization obeying the
condition
\begin{equation}
  \frac{1}{\beta} \artanh (m_i) -g(x_{l,j}) = x_{r,k}
\end{equation}
where $x_{l,j}$ and $x_{r,k}$ are points of the new natural partition of
$\msu{x}{l}$ and $\msu{x}{r}$ respectively. This condition leads to
\begin{equation}
  m_i =
\tanh \beta (f_{\{\sigma\}_l}(x^*_{\pm})+
A(f_{\{\tilde{\sigma}\}_r}(x^{*}_{\pm}))).
\end{equation}
We will employ these points as a new natural partition for the measure of the
local magnetization. It turns out that there exists a natural degeneracy
 within the set $\{m_i\}$ induced by the trivial identity
\begin{equation}
  A(a)+f_-(b) = A(b)+f_-(a)
\end{equation}
and other such identities comprising higher iterations of $f_+$ and $f_-$.
These degeneracies have to be removed ``by hand'' by the algorithm. In the
spirit of \cite{Halsey} we then define the partition function
\begin{equation}
  Z_{l,r} (q, \tau_q) = \sum_i \frac{\msu{m}{l,r}([m_{i+1}, m_i])^q}{(m_{i+1}
  - m_i)^{\tau_q}} 
\end{equation}
on this new natural partition and determine $D_q$ from the condition
\begin{equation}
  \ln Z_{l,r} (q, \tau_q) - \ln Z_{l', r'} (q, \tau_q) \stackrel{!}{=} 0 
\end{equation}
with some iteration depths $l$, $r$, $l'$, $r'$.

\begin{figure}
\psfrag{0}{\raisebox{-1mm}{$0$}}
\psfrag{1}{\raisebox{-1mm}{$1$}}
\psfrag{2}{\raisebox{-1mm}{$2$}}
\psfrag{3}{\raisebox{-1mm}{$3$}}
\psfrag{4}{\raisebox{-1mm}{$4$}}
\psfrag{5}{\raisebox{-1mm}{$5$}}
\psfrag{0.5}{\raisebox{-1mm}{$0.5$}}
\psfrag{1.5}{\raisebox{-1mm}{$1.5$}}
\psfrag{hc3}{\raisebox{-1mm}{$h_c^{(m,3)}$}}
\psfrag{hc4}{\raisebox{-1mm}{$h_c^{(m,4)}$}}
\psfrag{a}{a)}
\psfrag{b}{b)}
\psfrag{c}{c)}
\epsfig{file=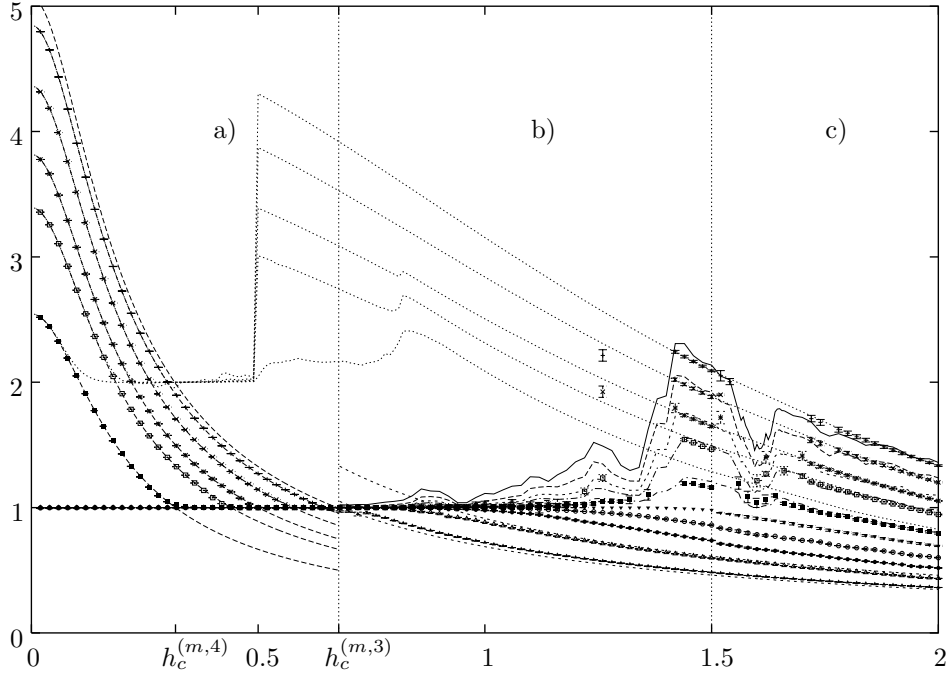, width=\textwidth}
\caption{Numerical results for the $D_q$-spectrum of the measure of the local
  magnetization in the bulk.  We considered $q= -20$, $-6$, $-3$, $-2$, $-1$,
  $0$, $1$, $2$, $4$ and $20$. All points with errorbars were obtained by the
  method based on the new natural partition. In regions a) and b) and $q
  \not= 0$ the results
  of all combinations of iteration depths $l=r=8$, $l=r=9$ and $l=r=10$ were
  used and the number representation was C++ long doubles. In region c) and
  also $q \not= 0$ all
  combinations of iteration depths $l=r=5$, $l=r=6$ and $l=r=7$ were used
  with numbers of the arbitrary precision library ``CLN'' with guaranteed
  $50$ decimal digits. The errorbars are obtained from the standard
  deviations of the average of the results of the three possible combinations
  of iteration depths. All points with standard deviation greater than $0.05$
  are not shown.
  For $q=0$ we used iteration depths up to $l=r=13$.
  The dashed lines in a) are lower bounds on $D_q$ and the
  dashed lines in b) and c) are upper bounds both based on the pointwise
  dimension at the boundary of the support of $\mu^{(m)}$. The dotted lines
  are the twofold numerical results for the $D_q$ ($q < 0$) of the effective
  field (cf \cite{Nowotny}) and are thus upper bounds on the $D_q$ ($q< 0$).
  Finally, the other lines in b) and c) are the results for $D_q$ ($q < 0$)
  obtained by the box scaling approach. The usual spacing in $h$ for all data
  points is $0.02$ except for the region between $h=0.56$ and $h=0.66$ where
  we chose a finer spacing of $\Delta h= 0.005$ in the box method. $\beta=J=1$.
  \label{fig7} }
\end{figure}

%%% Local Variables: 
%%% mode: latex
%%% TeX-master: article
%%% TeX-master: "article"
%%% End: 
 Figure \ref{fig7} shows a summary of the obtained numerical
results. In region a) the behaviour of the $D_q$ with negative $q$ is
dominated by the pointwise dimension at the boundary of the support of
$\mu^{(m)}$ such that the lower bounds based on it coincide with the obtained
numerical values. Furthermore, the numerical results are very stable for all
iteration depths. We do not show the result of the box method in this region
as it is known that box methods systematically underestimate $D_q$ for $q <
0$ if it strongly depends on only few points in the support. The generalized
dimensions for $q > 0$ are all $1$ because the measure is smooth. The
numerical results are in perfect agreement with this statement.

In region b) all $D_q$ for $q < 0$ are $1$ for $h \gtrsim h_c^{(m,3)}$.
At some point they again are greater than $1$. In this region the method
based on the new natural partition yields completely different results for
different iteration depths. We therefore are not able to deduce the
asymptotic behaviour from the scaling in finite iteration depth. Most data
points had to be left out for this reason. Provided $h$ is large enough
($h \gtrsim 1.7$) the numerical results of the new natural partition
method are again stable for all iteration depths which leads to small error
bars in figure \ref{fig7}. For $q > 0$ we have in regions b) and c) perfect
agreement with the upper bounds obtained from the pointwise dimension at the
boundary of the support of $\mu^{(m)}$.

The difficulties in obtaining the asymptotic scaling for $q < 0$ in the
region $1 \lesssim h \lesssim 1.7$ are of the same type as encountered
in the convolution of two-scale Cantor sets. This shows that this is an
effect of more than one relevant scale present (infinitely many in this case).
As the asymptotic scaling seems not to be attainable we have the impression
that from a physicists point of view we should pose the question what an
experimentalist would observe. In any experiment the scale of resolution is
bounded from below. This corresponds to the situation of the box method where
the scale is bounded by the size of the smallest box (whereas the scale in
the new natural partition can get more or less arbitrarily small already for
finite iteration depths).  We therefore surmise that the $D_q$ estimates
based on the box method are the physical results in this region. As it turns
out the results of the box method are fairly robust against changes in the
iteration depth whereas they depend on an appropriate choice of box sizes. As
a rule of thumb the smallest box size should be of the order of the length of
the longest band of the new natural partition. The results of the box method
are shown as lines in regions b) and c) in figure \ref{fig7}.

\begin{figure}
\begin{indented}
  \item
    \figwidth=\textwidth
    \addtolength{\figwidth}{-\mathindent}
    \unitlength\figwidth
    \begin{picture}(0,0.7)
      \psfrag{-104}{\fnt $10^{4}$}
      \psfrag{-102}{\fnt $10^{2}$}
      \psfrag{-1}{\fnt $\,1$}
      \psfrag{0}{\fnt $0$}
      \psfrag{1}{\fnt $1$}
      \psfrag{s0}{\tiny $\pts{5} 0$}
      \psfrag{102}{\fnt $\!10^{2}$}
      \psfrag{104}{\fnt $\!10^{4}$}
      \psfrag{-3}{\fnt $\!\!-3$}
      \psfrag{-2}{\fnt $\!\!-2$}
      \psfrag{xmp}{\fnt $\!x^*_{\{-+\}}$}
      \psfrag{sxmp}{\raisebox{2mm}{\tiny $\pts{2} x^*_{\{-+\}}$}}
      \psfrag{xpm}{\fnt $\!x^*_{\{+-\}}$}
      \psfrag{2}{\fnt $\!2$}
      \psfrag{3}{\fnt $\!3$}
      \psfrag{xlabel}{\fnt \raisebox{-1mm}{$x$}}
      \psfrag{ylabel}{\fnt \raisebox{1mm}{$\nts{45} A_\# \rho^{(x)}_r(y-x)$
        $\pts{20}\rho^{(x)}_l(x)$}}
      \put(0,0){\epsfig{file=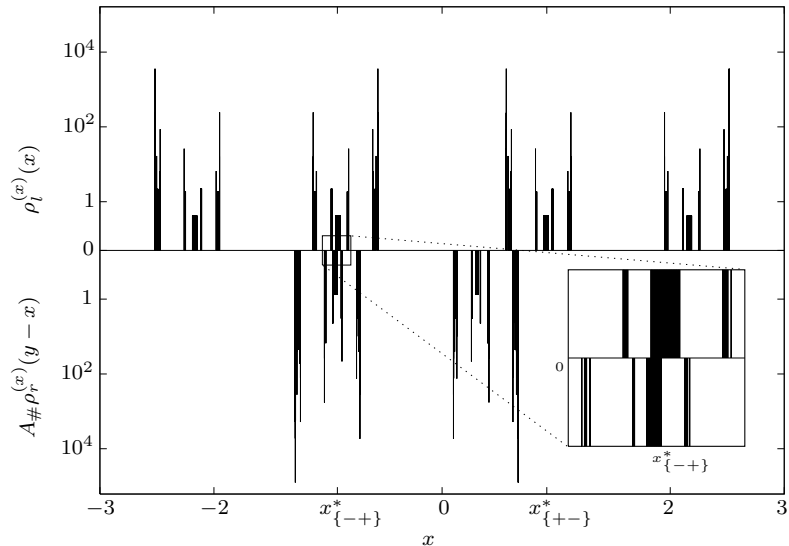, width=\figwidth}}
    \end{picture}
\end{indented}
\caption{Illustration of the situation leading to the weakest band in the
  convolution of $\mu^{(x)}_l$ and $A_\# \mu^{(x)}_r$ at $h=1.54$ and
  $l=r=6$. The integration over the product of $\rho^{(x)}_l$ (the density of
  $\smash \mu^{(x)}_l(x)$, upper part) and $\smash A_\# \rho^{(x)}_r(y-x)$
  (the density of $\smash A_\# \mu^{(x)}_r$, lower part) yields the density
  of the convolution at $y$. In this figure $y=0.309965$ which is the
  position of the weakest band. Only for the weak band around
  $x^*_{\{-+\}}$ the two densities are simultaneously non-zero (see inlay). This leads to
  a very small value for the density of the convolution. Note that
  even though the two convoluted measures are already rather sparse, the
  convolution still has non-fractal support at this $h$. $\beta=J=1$.
  \label{fig8}}
\end{figure}

The local minimum of the $D_q$, $ q < 0$ at $h \approx 1.6$ can be understood
as a change in the overlap structure of bands of the new natural partition of
the two convoluted measures for different $h$. For $h \approx 1.5$ there is a
relative position for the approximations $\mu^{(x)}_l$ and $A_\# \mu^{(x)}_r$
for which only the two very weak bands around $x^*_{+-}$ and $A(x^*_{+-})$
overlap leading to a very weak band in the convolution, cf figure \ref{fig8}
for an example. This results in the
large values of $D_q$ we observe numerically for the iteration depths $l=r
\leq 10$. For $h \approx 1.6$ the band structure is of a form that no such
position can be found in the iteration depths under consideration. For any
relative position of the two approximations of the measures more than one
pair of bands or considerably stronger ones overlap. This results in the
considerably smaller values of $D_q$. For larger $h$ the formation of a very
weak band in the convolution reappears and we again get large values of $D_q$
($q < 0$). For other iteration depths the situation can again change as there
is no strict self-similarity of the measure $\mu^{(m)}$.
From another point of view for a given iteration depth the values of $D_q$
strongly depend on the random field strength $h$ due to similar changes in
the overlap structure as discussed so far. We expect that on
any iteration depth this situation is qualitatively the same.

%%% Local Variables:
%%% mode: latex
%%% TeX-master: article
%%% End: 

\begin{figure}
  \setlength{\figwidth}{\textwidth}
  \addtolength{\figwidth}{-1cm}
  \psfrag{0}{\raisebox{-0.5mm}{$\!\sst 0$}}
  \psfrag{m}{\raisebox{-0.5mm}{$\!\!\!\sst 12$}}
  \psfrag{-0.1}{\raisebox{-0.5mm}{$\!\!\!\sst -0.1$}}
  \psfrag{0.1}{\raisebox{-0.5mm}{$\!\!\sst 0.1$}}
  \psfrag{h}{$\nts{5} \sst h=0.02$}
  \mbox{} \hspace{0.7cm}
\epsfig{file= 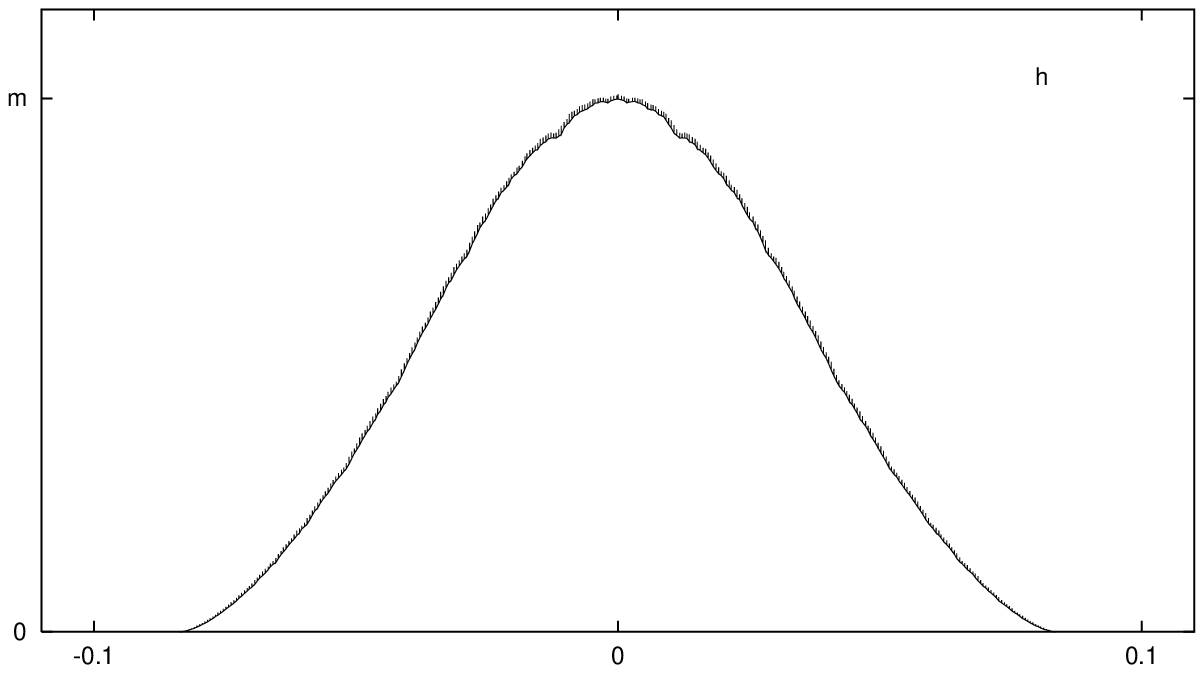, width= 0.45\figwidth}
  \psfrag{-1}{\raisebox{-0.5mm}{$\!\!\sst -1$}}
  \psfrag{1}{\raisebox{-0.5mm}{$\!\sst 1$}}
  \psfrag{m}{\raisebox{-0.5mm}{$\!\!\!\sst 10$}}
\hfill \epsfig{file= 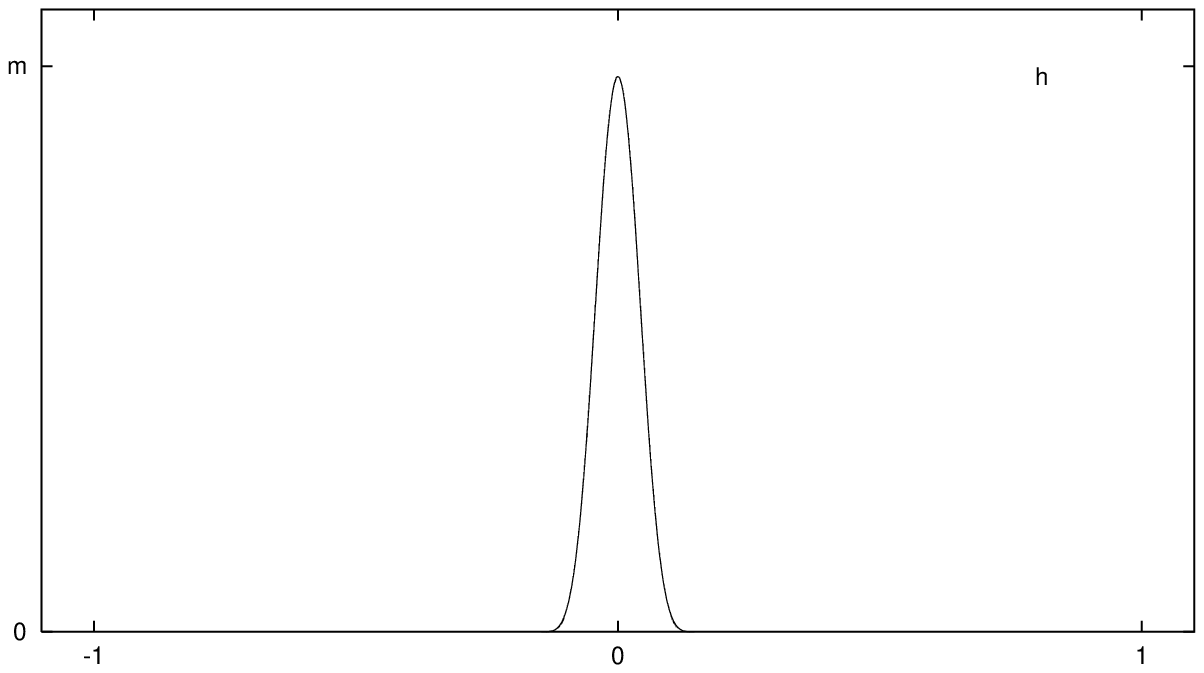, width= 0.45\figwidth} \\[0.18cm]
  \psfrag{m}{\raisebox{-0.5mm}{$\!\!\sst 1$}}
  \psfrag{h}{$\nts{3} \sst h=0.2$}
  \psfrag{-0.8}{\raisebox{-0.5mm}{$\nts{3} \sst -0.8$}}
  \psfrag{0.8}{\raisebox{-0.5mm}{$\nts{2} \sst 0.8$}}
  \mbox{} \hspace{0.7cm}
\epsfig{file= 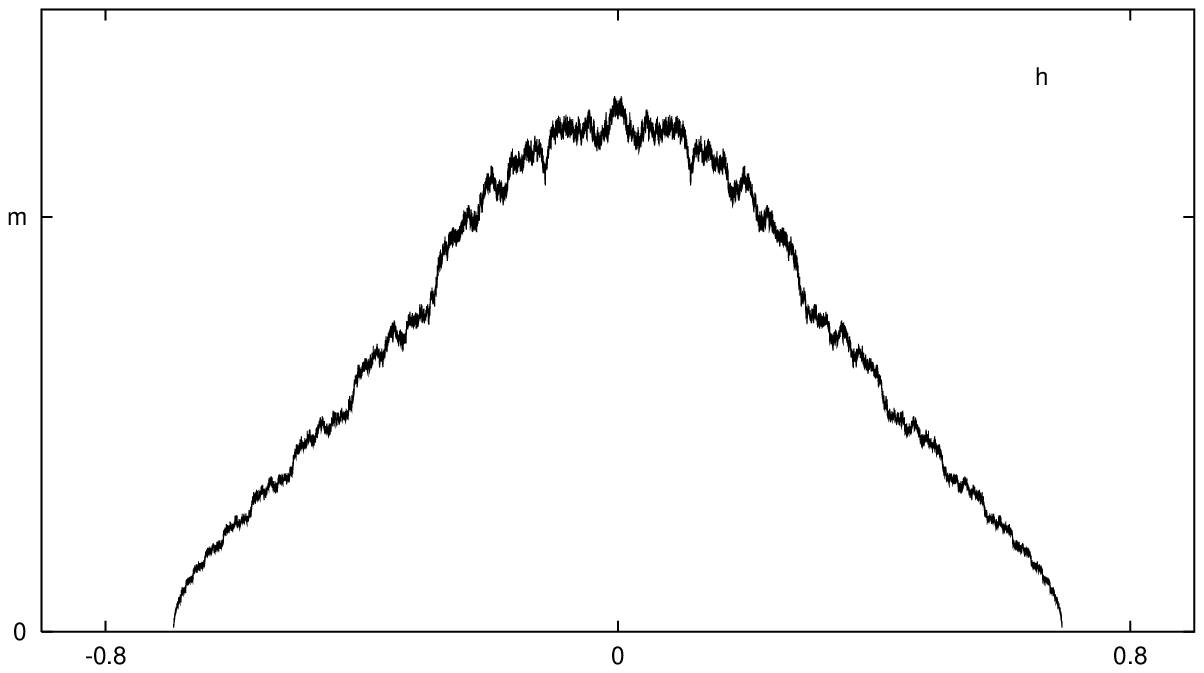, width= 0.45\figwidth}
  \psfrag{m}{\raisebox{-0.5mm}{$\!\!\sst 1$}}
\hfill \epsfig{file= 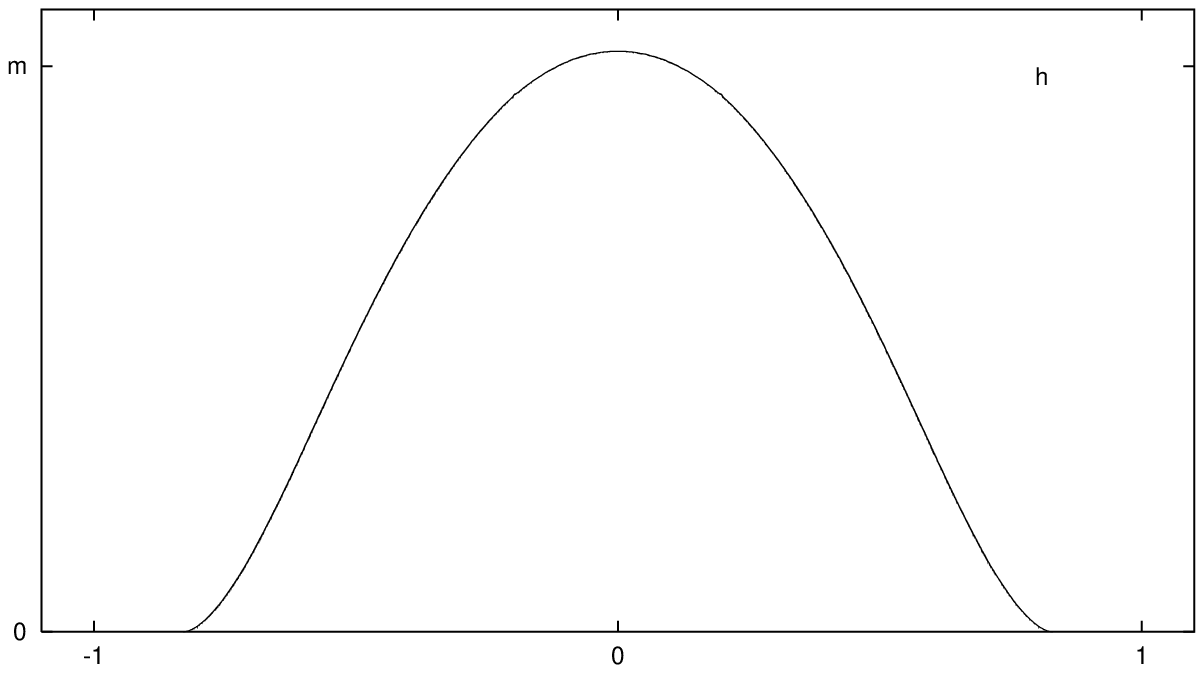, width= 0.45\figwidth} \\[0.18cm]
  \psfrag{m}{\raisebox{-0.5mm}{$\!\!\sst 2$}}
  \psfrag{h}{$\nts{3} \sst h=0.4$}
  \mbox{} \hspace{0.7cm}
\epsfig{file= 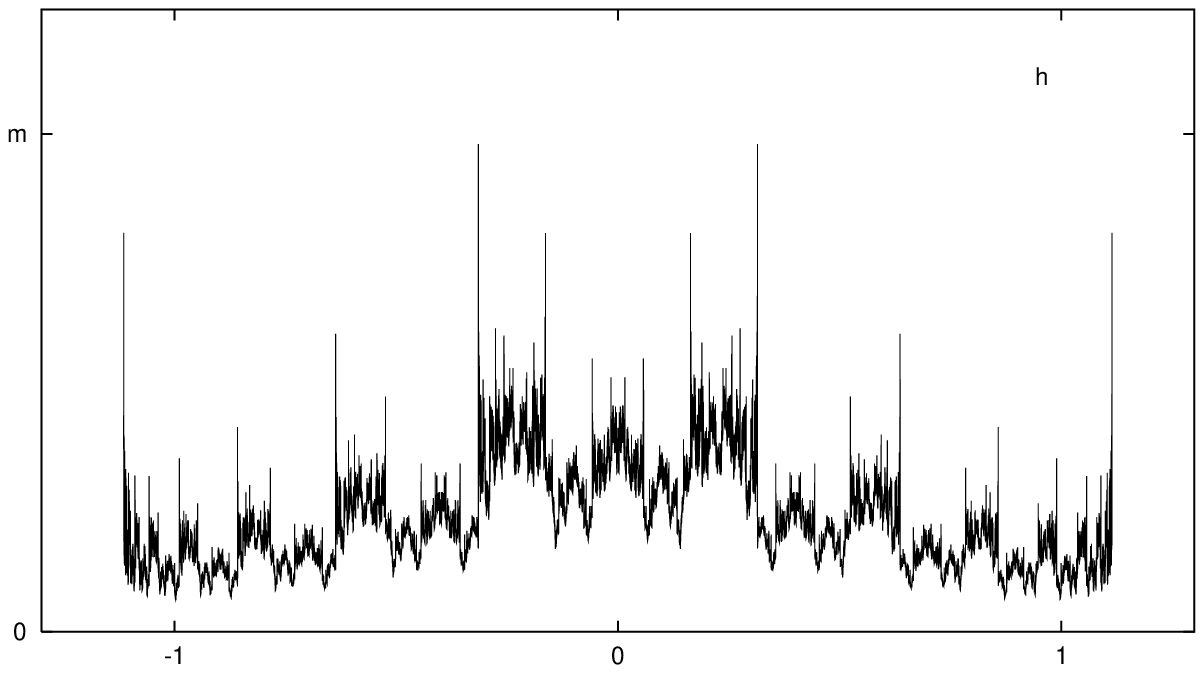, width= 0.45\figwidth}
  \psfrag{m}{\raisebox{-0.5mm}{$\!\!\!\!\!\sst 0.7$}}
\hfill \epsfig{file= 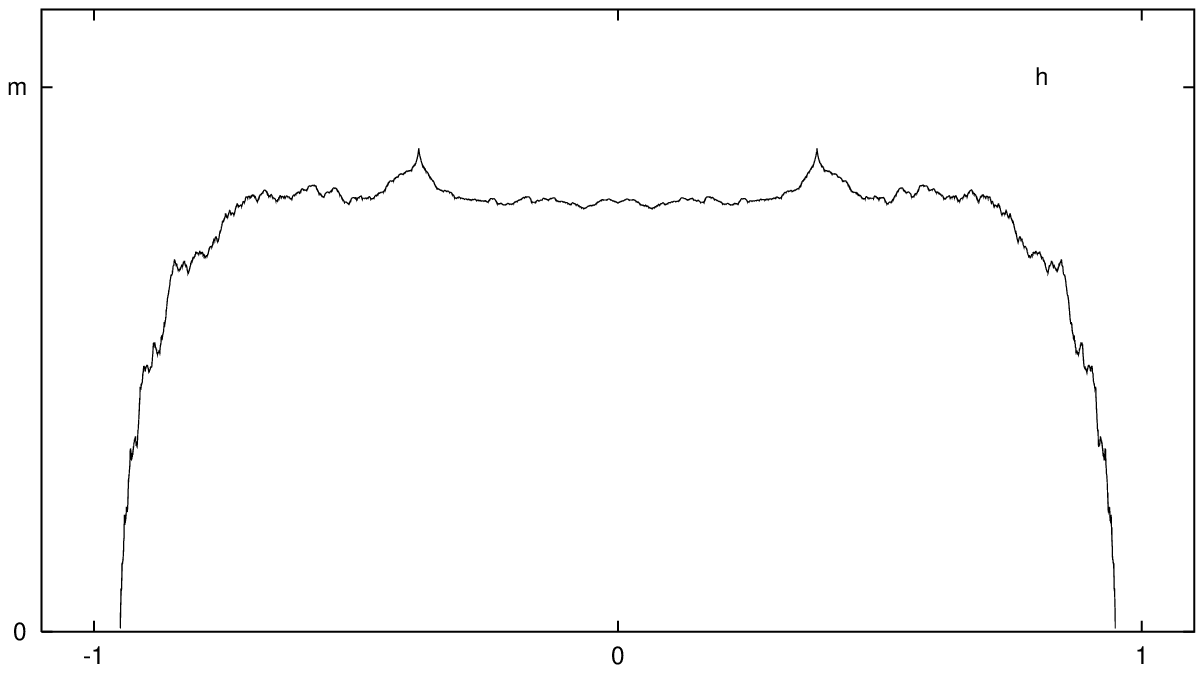, width= 0.45\figwidth} \\[0.18cm]
  \psfrag{m}{\raisebox{-0.5mm}{$\nts{7}\sst 3000$}}
  \psfrag{h}{$\nts{3} \sst h=0.7$}
  \psfrag{-1.5}{\raisebox{-0.5mm}{$\nts{3} \sst -1.5$}}
  \psfrag{1.5}{\raisebox{-0.5mm}{$\nts{2} \sst 1.5$}}
  \mbox{} \hspace{0.7cm}
\epsfig{file= 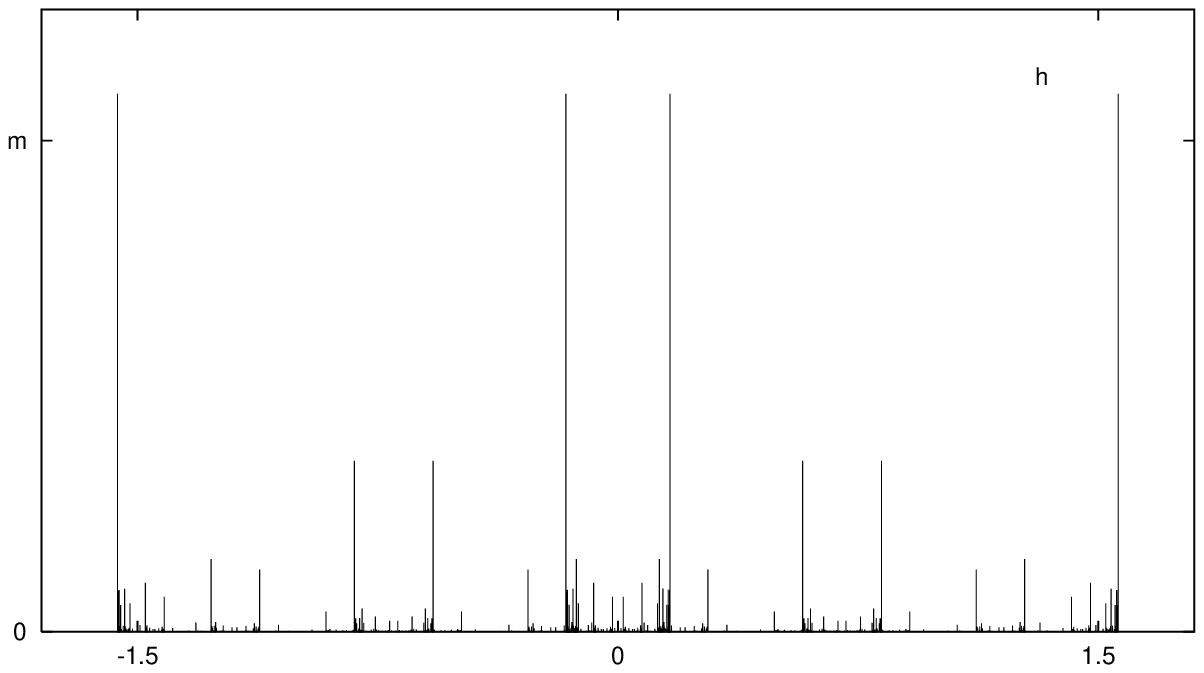, width= 0.45\figwidth}
  \psfrag{m}{\raisebox{-0.5mm}{$\!\!\!\!\!\sst 2.5$}}
\hfill \epsfig{file= 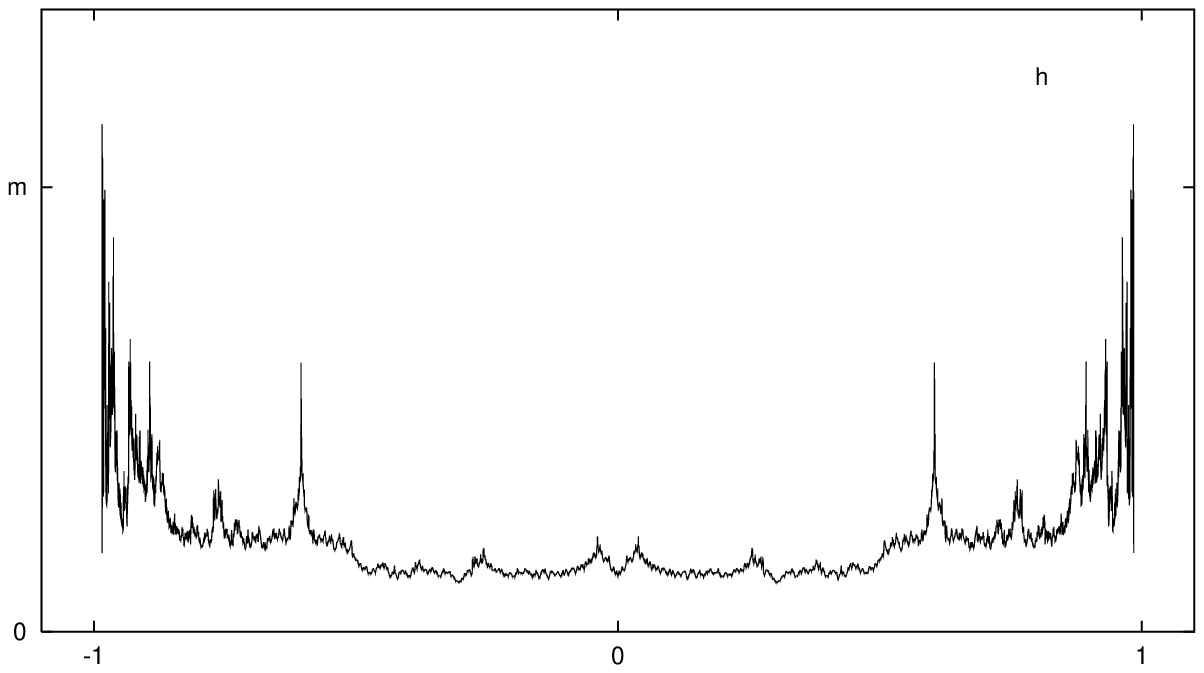, width= 0.45\figwidth}  \\[0.17cm]
  \psfrag{m}{\raisebox{-0.5mm}{$\nts{6} \sst 10^{7}$}}
  \psfrag{h}{$\sst h=1$}
  \mbox{} \hspace{0.7cm}
\epsfig{file= 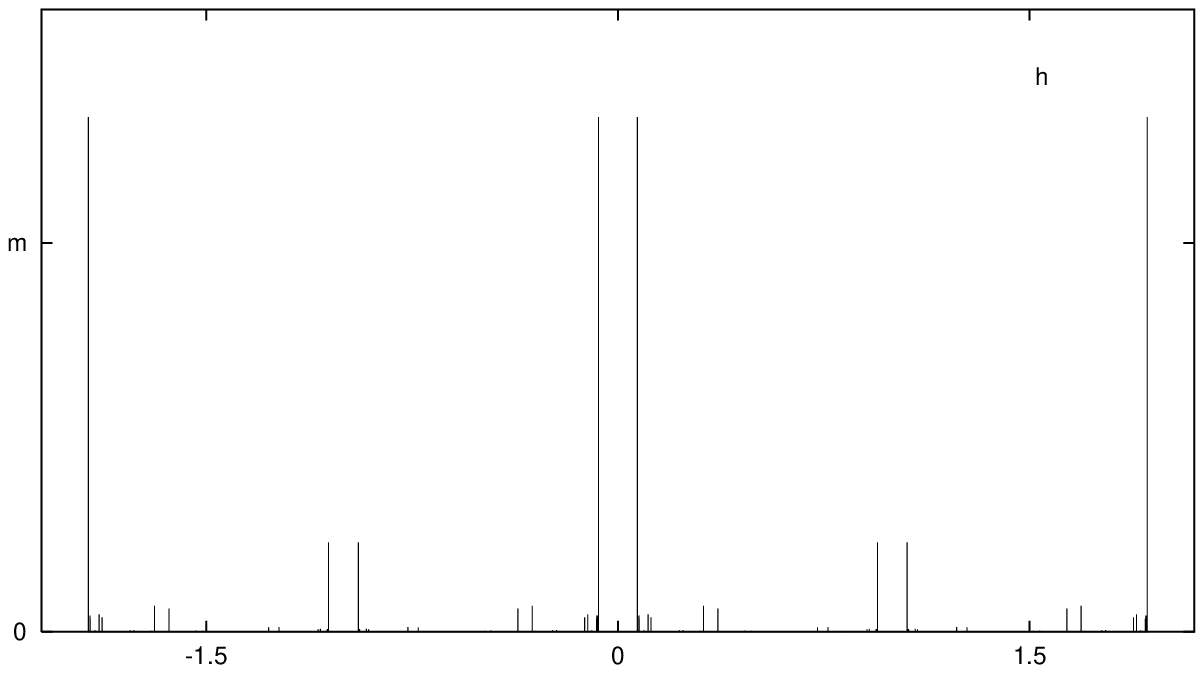, width= 0.45\figwidth}
  \psfrag{m}{\raisebox{-0.5mm}{$\nts{6} \sst 150$}}
\hfill \epsfig{file= 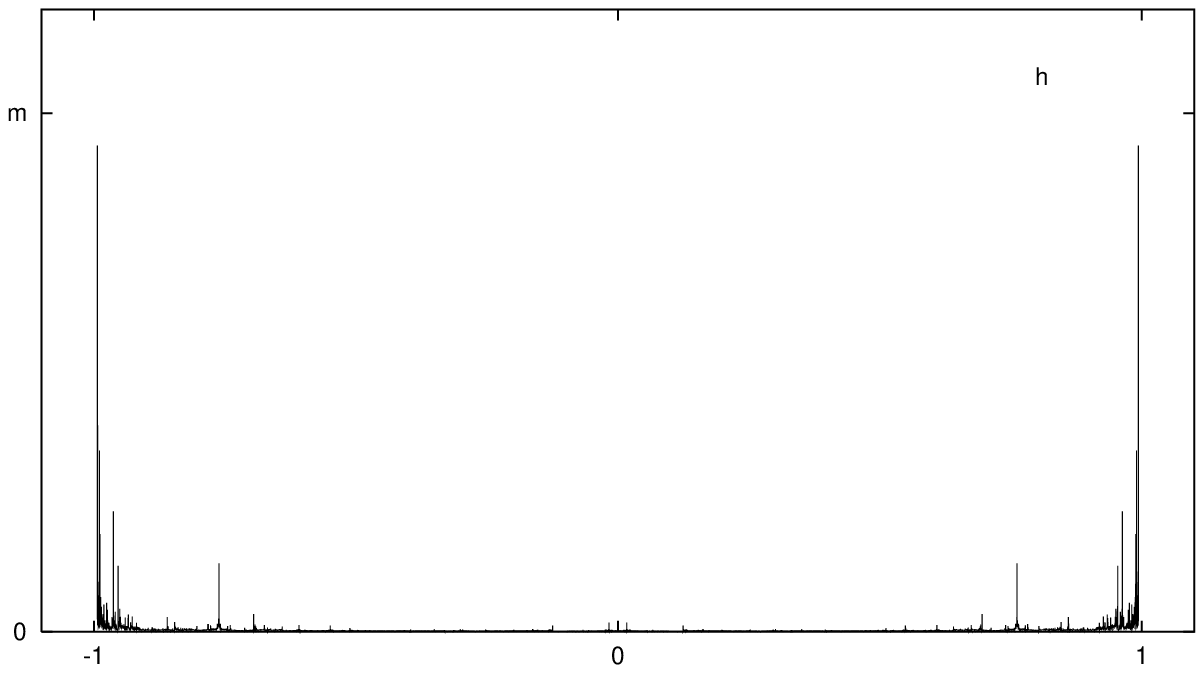, width= 0.45\figwidth} \\[0.17cm] 
  \psfrag{m}{\raisebox{-0.5mm}{$\nts{8} \sst 10^{11}$}}
  \psfrag{h}{$\nts{3} \sst h=1.3$}
  \psfrag{-2}{\raisebox{-0.5mm}{$\nts{2} \sst -2$}}
  \psfrag{2}{\raisebox{-0.5mm}{$\nts{1} \sst 2$}}
  \mbox{} \hspace{0.7cm}
\epsfig{file= 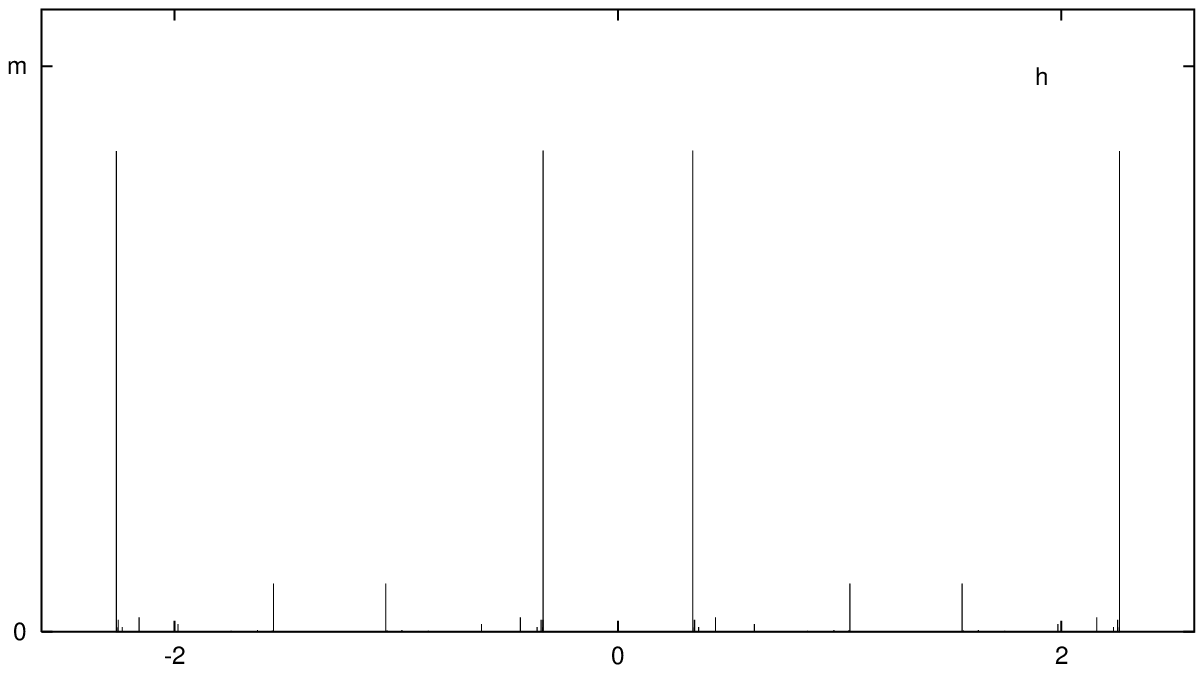, width= 0.45\figwidth}
  \psfrag{m}{\raisebox{-0.5mm}{$\nts{10}\sst 15000$}}
\hfill \epsfig{file= 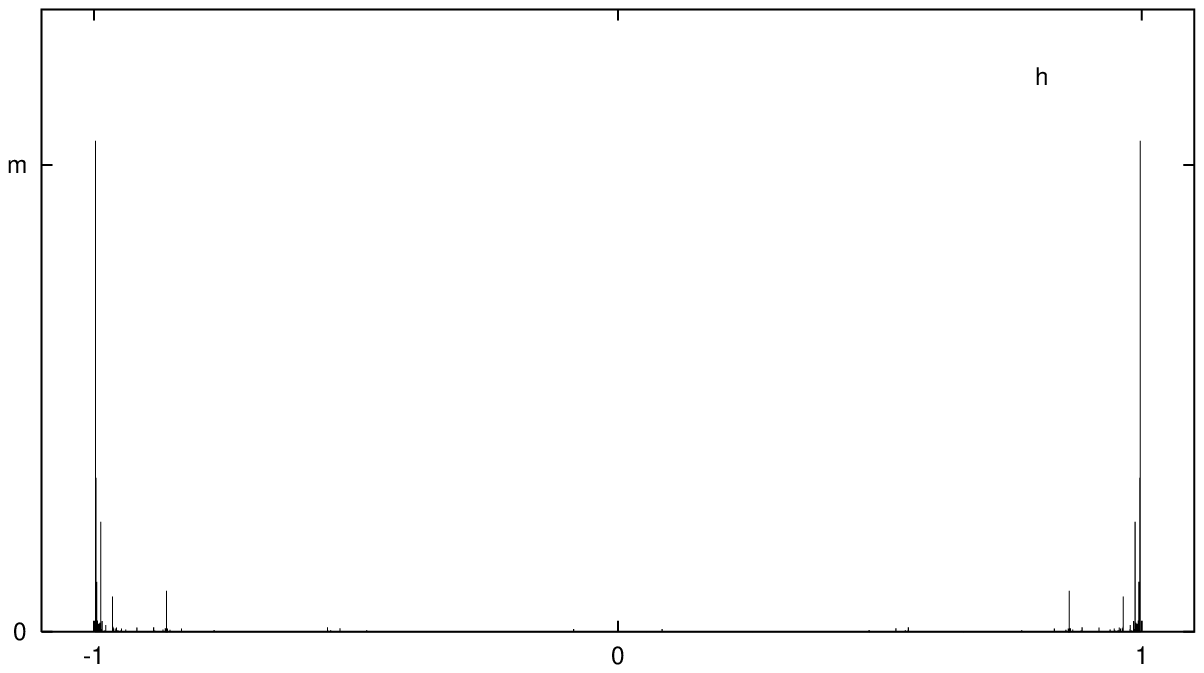, width= 0.45\figwidth}
\caption{Measure densities of the effective field (left column) for iteration
  depth $16$ and the local magnetization (right column) for iteration depth
  $l=r=8$. The details are explained in the text. $\beta= J= 1$.\label{fig9}}
\end{figure}

%%% Local Variables: 
%%% mode: latex
%%% TeX-master: article
%%% End: 

\section{Conclusions} \label{concludesec}
In order to investigate the multifractal properties of the measure of the
local magnetization in the random-field Ising chain we studied the relation
between the multifractal properties of two measures and their convolution.
The pointwise dimension at the boundary of the support of the convolution
turned out to be the sum of the pointwise dimensions at the boundary of the
two measures. This enabled us to calculate the pointwise dimension at the
boundary of the support of the measure of
the local magnetization and employ this to give bounds on the $D_q$-spectrum.

We furthermore were able to prove that the generalized dimensions $D_q$ of
the convolution are bounded from above by the sum of the $D_q$ of the two
convoluted measures. This yields upper bounds on the $D_q$-spectrum of the
measure of the local magnetization.

The general results were illustrated for the convolution of Cantor sets with
weights and our main application, the measure of the local magnetization in
the random-field Ising model. We also performed numerical studies of the
$D_q$-spectra employing a box method and a method based on a new natural
partition. The numerical data is consistent with the exact inequalities but
reveals a complicated structure for the $D_q$-spectrum of the measure of the
local magnetization for $q<0$ and $1.2 \lesssim h \lesssim 1.7$. The
numerical instability of the method based on the new natural partition in
this region can be understood by analysing the band structure of the
approximations of the invariant measure of the effective field. Nevertheless,
it prevents us from obtaining the asymptotic scaling.  We therefore took a
pragmatic approach and investigated what an experimentalist could possibly
hope to observe resulting in the shown $D_q$ estimates based on the box
method.

Investigating the probability measure of the local magnetization of the
one-dimensional random-field Ising model we for the first time considered a
physical quantity which in principle is measurable. Furthermore, the general
results about convolutions of multifractal measures can be of interest in
other areas as well as it is not unusual that sums of random variables
appear.

All figures were presented for a generic choice of parameters, $\beta=J=1$.
The behaviour for $\beta, J \to 0$ and $\beta, J \to \infty$ is more or less
trivial. For $\beta \to 0$ the probability measure of the local magnetization
is a Dirac measure at $0$. For $J \to 0$ it is the sum of two Dirac measures
at $\pm \tanh \beta h$. In the limit $\beta \to \infty$ the distribution of
the effective field is a finite sum of Dirac measures \cite{Behn1,Behn4} and
therefore the measure of the local magnetization is also such a sum. Finally,
in the unphysical limit $J \to \infty$ the function $A(x)$ is the identity,
$A(x)=x$, and the RIFS for the effective field therefore ceases to be
contractive. The measures $\mu_n$ of the effective field for finite system
sizes need not converge in Hutchinson metric in the thermodynamic limit $n
\to \infty$. The proof for existence and uniqueness of the convolution of
lemma \ref{lemma1} does therefore not apply. One needs to consider finite
systems and take the thermodynamic limit in the end. The RIFS is a symmetric
random walk with step size $h$ and the local magnetization is $m_{l,r} =
\tanh \beta (x_l + y_r)$ where $x_l$ and $y_r$ are two random walks of length
$l$ and $r$ respectively. Thus, $1/\beta \, \artanh m_{l,r}$ is also a
random walk of length $n=l+r$. One therefore can deduce that the probability
for the local magnetization to take values in any closed interval $X \subset
[-1,1]$ tends to zero in the thermodynamic limit $n \to \infty$ because this
corresponds to the probability for the random walk to stay in a finite
region. The measure of the local magnetization thus converges to the sum of
two Dirac measures at $\pm 1$ in the weak topology of Borel measures. In all
cases no multifractal effects of interest can be observed.

As a byproduct of our numerical algorithm we can easily produce the measure
density of the local magnetization itself, cf figure \ref{fig9}. In the
figure we show the measure density $\rho^{(x)}$ of the effective field and
the measure density $\rho^{(m)}$ of the local magnetization for some values
of the random field strength $h$. One can clearly see that $\rho^{(m)}$ is
much smoother than $\rho^{(x)}$ in accordance with the general belief. For
$h= 0.02$ both measures are smooth and the slope at the boundary is zero.
For $h= 0.2 > h_c^{(4)}$ the slope of $\rho^{(x)}$ is already infinite
whereas the slope of $\rho^{(m)}$ remains zero. For $h= 0.4 > h_c^{(3)} =
h_c^{(m,4)}$ the density $\rho^{(x)}$ is infinite at the boundary whereas
$\rho^{(m)}$ is zero but has infinite slope. For $h= 0.7 > h_c^{(m,3)}$ the
density $\rho^{(m)}$ also is infinite at the boundary. The fractality of the
support is in the same way ``delayed'' for $\rho^{(m)}$. For $h=1.0 >
h_c^{(1)}$ the support of $\rho^{(x)}$ is already fractal but the support of
$\rho^{(m)}$ is still Euclidean.

Overall, there is a gradual transition from a monomodal strongly peaked
distribution for small random field to an even more strongly peaked bimodal
distribution for large random field.  The local magnetization (which is a
ther\-mo\-dy\-namic average but still a random variable with respect to the
probability space of the random field) shows a transition from a paramagnetic
situation where the most probable value is zero to a ferromagnetic situation
where the most probable value is $\pm 1$.  Between these extremal situations
lies the multifractal regime. The distribution always remains symmetric such
that this is not a phase transition; there is no symmetry breaking even if a
small homogeneous field is applied.

\begin{appendix}
\section{} \label{appe1}
  Let $x_i= i\eps$, $I \in \Z$ and $x_i'= x_i+y$, $i \in \Z$ be two grids
  which are shifted by $y$ with respect to each other and let $q > 0$. In this
  appendix we show that
  \begin{equation}
    \sum_{i} \mu(B_{\frac{\eps}{2}} (x_i'))^q \leq 2^{q+1} \sum_{i}
    \mu(B_{\frac{\eps }{2}}(x_i))^q. \label{ineq4}
  \end{equation}
  Let $i \in \Z$ and denote $\mu_i:= \mu(B_{\frac{\eps}{2}}(x_i))$ and
  $\mu_i':= \mu(B_{\frac{\eps}{2}}(x_i'))$. Clearly,
  $\mu_i' \leq \sum_{j \in J(i)} \mu_j$
  where $J(i)= \{ j \in \Z: B_{\frac{\eps}{2}}(x_j) \cap
  B_{\frac{\eps}{2}}(x_i') \not= \emptyset\}$. As $B_{\frac{\eps}{2}}(x_i')$
  intersects at most two $B_{\frac{\eps}{2}}(x_j)$ the set $J(i)$ has at most
  two elements and we can write
  \begin{equation}
    {\mu_i'}^q \leq 2^q \max_{j \in J(i)} \mu_j^q. \label{ineq3}
  \end{equation}
  On the other hand each $B_{\frac{\eps}{2}}(x_j)$ intersect at most two
  $B_{\frac{\eps}{2}}(x_i)$ for a fixed $j$ such that $\mu_j$ appears at most
  twice when summing (\ref{ineq3}) over all $i$. Therefore,
  \begin{equation}
    \sum_i {\mu_i'}^q \leq 2\cdot 2^q \sum_i \mu_i^q
  \end{equation}
  which is the result claimed in (\ref{ineq4}).
\end{appendix}

\end{document}